\documentclass[10pt]{article}
\usepackage{amsfonts}
\usepackage{color,caption}
\usepackage{graphicx}
\usepackage[export]{adjustbox}
\usepackage{amsmath}
\usepackage{float}
\usepackage{mathtools}
\usepackage{amssymb}
\providecommand{\U}[1]{\protect\rule{.1in}{.1in}}
\textwidth=6.5in \hoffset=-.75in \voffset=-1in \numberwithin{equation}{section}
\numberwithin{figure}{section}

\newcommand {\be}{\begin{equation}}
	\newcommand {\ee}{\end{equation}}
\newcommand {\bea}{\begin{eqnarray}}
	
	\newcommand {\eea}{\end{eqnarray}}
\newcommand {\tcr}{\textcolor{red}}

\newcommand{\rbm}[1]{{\color{red} \bf [Robb: #1]}}

\begin{document}
	\begin{titlepage}
		\bigskip \begin{flushright}
		\end{flushright}
		\vspace{1cm}
		\begin{center}
			{\Large \bf {Holographic Thermodynamics of Dyonic Dilaton AdS Black Holes}}\\
			\vskip 1cm
		\end{center}
		\vspace{1cm}
				\begin{center}
			Muhammad Fitrah Alfian Rangga Sakti$^{a,b,c,}${\footnote{fitrahalfian@gmail.com}}, Robert Mann$^{b,c,}${\footnote{rbmann@uwaterloo.ca }} and Piyabut Burikham$^{a,}${\footnote{piyabut@gmail.com}}
			\\
			$^a $High Energy Physics Theory Group, Department of Physics,
			Faculty of Science, Chulalongkorn University, Bangkok 10330, Thailand\\
			$^b $Department of Physics and Astronomy, University of Waterloo, Waterloo, Ontario, N2L 3G1, Canada\\
			$^c $Perimeter Institute for Theoretical Physics, Waterloo, Ontario, N2L 2Y5, Canada \\
			\vspace{1cm}
		\end{center}
		
		\begin{abstract}
We investigate the holographic dual of the extended bulk thermodynamics of dyonic dilaton Anti-de Sitter (AdS) black holes by allowing both the cosmological constant and Newton's gravitational constant to vary. In the dual CFT thermodynamics, the central charge $C$ and chemical potential $\mu$ as its conjugate enter the thermodynamic relations, in addition to temperature-entropy $(\tilde{T},\tilde{S})$, electric charge-potential $(\tilde{Q},\tilde{\Phi})$, magnetic charge-potential $(\tilde{P},\tilde{\Psi})$, pressure-volume $(\tilde{\mathcal{P}},\tilde{V})$.
We consider sixteen different ensembles related to those five pairs of thermodynamic variables which we separate into fixed volume and pressure ensembles. Due to the symmetry on ($\tilde{Q},\tilde{P}$), we only need to analyze five ensembles each in fixed volume and pressure ensembles. In the fixed volume ensembles, it is found that the fixed ($\tilde{Q},\tilde{P},\tilde{V},\mu$) and ($\tilde{\Phi},\tilde{P},\tilde{V},\mu$) exhibit a zeroth-order phase transition. Interestingly, the fixed ($\tilde{\Phi},\tilde{\Psi},\tilde{V},C$) ensemble shows the zeroth- and first-order phase transitions at above and below critical potential $\tilde{\Upsilon}_c$, respectively where $\tilde{\Upsilon}^2=\tilde{\Phi}^2+\tilde{\Psi}^2$. In the fixed pressure ensembles, it is found that richer structures appear where the sign of $\mu$ heavily influences phase space for several ensembles. For fixed ($\tilde{Q},\tilde{P},\tilde{\mathcal{P}},\mu$) and ($\tilde{\Phi},\tilde{P},\tilde{\mathcal{P}},\mu$), they show the zeroth-, first-, and second-order phase transitions. For the fixed ($\tilde{\Phi},\tilde{\Psi},\tilde{\mathcal{P}},\mu$) ensemble, there is only a zeroth-order phase transition occurs. The other ensembles which we do not mention eventually do not show any phase transition. These findings provide deeper insight into how bulk gravitational variations, particularly regarding the cosmological and gravitational constants, translate to exact critical phenomena and richer phase structures within conformal field theories.
		\end{abstract}
	\end{titlepage}\onecolumn
	\bigskip
	
\section{Introduction}
\label{sec:Intro}

Since Hawking’s seminal work on black hole radiation \cite{HawkingCMP1975,HawkingPRD11976}, the study of black hole thermodynamics has grown into one of the most attractive areas of research in gravitational physics. It is now widely viewed as a promising route toward understanding the quantum nature of gravity. Over the past few decades, the thermodynamic properties of many black hole solutions have been explored in detail, leading to several important developments. These include the connection between the laws of gravitation and the laws of thermodynamics \cite{JacobsonPRL1995,PadmanabhanRPP2010}, for instance, the study of black hole thermodynamics from a geometric perspective \cite{AmanBengtssonGRG2003,QuevedoGRG2008,RuppeinerSPP2014,MansooriMirzaEPJC2014,SureshTharanathEPJC2014,ZhangCaiJHEP2015,ZhangCaiPRD2015,HendiSyekhiPRD2015,GruberLuongoProc2015}, the use of black holes as holographic models in condensed matter physics \cite{HartnollKovtunPRB2007,HartnollHerzogPRL2008} and quantum chromodynamics \cite{KovtunPRL2005}.
	
Recently, several studies show that the thermodynamics of black holes in  asymptotic anti-de Sitter (AdS) spacetime are quite distinct from those in flat or de Sitter (dS) space. In particular, since the emergence of  ``black hole chemistry" \cite{KubiznakMannCJP2014,KubiznakMannCQG2017,Mann:2025xrb}, the study of black hole thermodynamics has undergone a significant paradigm shift that follows the recognition that the cosmological constant $\Lambda$ should be treated as a thermodynamic pressure 
in an extended thermodynamic phase space that includes
its conjugate volume 
\cite{TeitelboimPLB1985,Creighton:1995au,CaldarelliCQG2000,KastorRayCQG2009,DolanCQG2011,DolanCQG2011a,CveticGibbonsPRD2011}. This framework has revealed a rich landscape of critical phenomena in bulk gravity, such as first-order transition from radiation
to a black hole (or vice-versa)
\cite{HawkingPageCMP1983},  which is related to confinement (deconfinement) transition of the dual quark gluon plasma \cite{WittenATMP1998}, 
and can be understood as a solid-liquid phase transition \cite{KubiznakMannCJP2014}.
Further investigations have led to the discovery of 
Van der Waals-like phase transitions \cite{ChamblinEmparanPRD1999,ChamblinEmparanPRD1999a,CveticGubserJHEP1999,KubiznakMannJHEP2012},   superfluid phase transitions \cite{HennigarMannPRL2017}, re-entrant phase transitions \cite{AltamiranoKubiznakPRD2013,FrassinoKubiznakJHEP2014},  triple points \cite{AltamiroKubiznakCQG2014,WeiLiuPRD2014}, and multicritical points
\cite{Tavakoli:2022kmo,Wu:2022bdk,Wu:2022plw}.

Currently, the AdS/CFT correspondence provides a powerful dictionary for exploring the dual boundary Conformal Field Theory (CFT) thermodynamics. Recently, this correspondence has been extended by treating Newton’s gravitational constant $G$ as a dynamical thermodynamic variable in addition to the extended bulk thermodynamics \cite{KastorRayCQG2010,KastorRayJHEP2014,KarchRobinsonJHEP2015,SarkarVisserJHEP2020,VisserPRD2022,CongKubiznakPRL2021}, or alternatively varying the conformal factor of the boundary metric
\cite{AhmedCongPRL2023}.
This formulation was raised when one considers the mapping of bulk pressure and thermodynamic volume to CFT thermodynamics, since both are not dual to the pressure and spatial volume in CFT thermodynamics, respectively. Moreover, the interpretation of the bulk pressure seemed incompatible with the holography \cite{ZhangCaiJHEP2015,ZhangCaiPRD2015,KastorRayJHEP2014,JohnsonCQG2014,DolanJHEP2014,DolanEntropy2016}. Currently, this correspondence, known as holographic thermodynamics, has been successful in mapping the bulk thermodynamics of charged and charged rotating AdS black holes in Einstein-Maxwell system \cite{CongKubiznakJHEP2022,AhmedCongJHEP2023}. This formulation helps to correspond the bulk pressure and thermodynamic volume in bulk thermodynamics to the quantities in CFT thermodynamics. Concretely, varying the cosmological constant simultaneously with varying Newton's gravitational constant corresponds to the varying of the central charge $C$ or the number of colors $N$ in the dual gauge theory. To bridge the bulk and CFT thermodynamics, interestingly one can construct such mixed thermodynamics by allowing the central charge $C$ of the dual boundary theory to emerge as a thermodynamic quantity with a corresponding chemical potential $\mu_c$ as its conjugate \cite{GongJiangJHEP2023}. By applying the AdS/CFT dictionary, the standard first law of the bulk thermodynamics can be mapped to the first law of the CFT thermodynamics via mixed thermodynamics. Furthermore, the generalized Smarr formula for black holes can be shown also that this is dual to the Euler formula in the CFT \cite{KarchRobinsonJHEP2015,VisserPRD2022}. Not only in Einstein-Maxwell system, this formulation is succesfully applied also to beyond this theory such as in Born-Infeld theory \cite{KumarSenPRD2022,BaiSongEPJC2024} and Gauss-Bonnet theory \cite{KumarSenPRD2023,QuTaoNPB2023} to investigate the phase transitions.

Our purpose here is to   systematically investigate the holographic dual of the extended thermodynamics for dyonic dilaton AdS black holes by varying both the cosmological constant and Newton's gravitational constant. We find that presence of the dilaton and axion fields  provides a much richer structure to the phase phase compared to that of charged AdS black holes \cite{CongKubiznakJHEP2022}. We map bulk parameters to their dual boundary CFT variables which are temperature-entropy $(\tilde{T},\tilde{S})$, electric charge-potential $(\tilde{Q},\tilde{\Phi})$, magnetic charge-potential $(\tilde{P},\tilde{\Psi})$, pressure-volume $(\tilde{\mathcal{P}},\tilde{V})$, and central charge-chemical potential ($C,\mu$), enabling us to explore the thermodynamic phase space of sixteen different ensembles. Due to the symmetry on ($\tilde{Q},\tilde{P}$), we only need to analyze five ensembles each in fixed volume and pressure ensembles. We aim to explore the phase transitions and stability of the CFT states, which are
dual to dyonic dilaton AdS black holes, providing a new perspective on how bulk gravitational variations translate to exact critical phenomena in boundary field theories.

After the Introduction in Section \ref{sec:Intro}, we carry out the dyonic dilaton AdS black hole from its rotating version by setting $a=0$ in Section \ref{sec:dyonic}. In Section \ref{sec:massform}, we focus on the construction generalized mass/energy formulas of dyonic dilaton AdS black holes in extended thermodynamics, mixed thermodynamics, and CFT thermodynamics. In Section \ref{sec:ensemble_constantV}, we delve into the phase transitions of the CFT thermodynamics for fixed spatial volume ensembles. Afterward, we will explore phase behaviors of the CFT states in the fixed pressure ensembles in Section \ref{sec:ensemble_constantPr}. We then present the summary in the last section.

\section{Dyonic Dilaton AdS Black Hole}
\label{sec:dyonic}
In this section we will carry out the dilaton black hole solution with non-vanishing dyonic charge in asymptotically AdS spacetime. This is a non-rotating version of the dyonic Kerr-Sen-AdS black hole. These black hole solutions are the solutions to the following action \cite{WuWuWuYuPRD2021}
	\begin{eqnarray}
		\mathcal{L}&=&\sqrt{-g}\left[R_s - \frac{1}{2}(\partial\varphi)^2-\frac{1}{2}e^{2\varphi}(\partial\chi)^2-e^{-\varphi}F^2 + \sqrt{-g}\frac{4+e^{-\varphi}+e^{\varphi}(1+\chi^2)}{l^2} \right] +\frac{\chi}{2}\epsilon^{\mu\nu\rho\lambda}F_{\mu\nu}F_{\rho_\lambda}, \label{eq:Lagrangian}\
	\end{eqnarray}
	where $R_s$ is the Ricci scalar, $\varphi$ is the dilaton field, $\chi$ is the pseudoscalar axion field, and $F$ is the electromagnetic field. $\epsilon^{\mu\nu\rho\lambda}$ is the four-dimensional Levi-Civita antisymmetric tensor density. The last term in the Lagrangian is the term containing the coupling between electric and magnetic sectors. In order to obtain the dyonic dilaton AdS black hole solution, we can start with the dyonic Kerr-Sen-AdS black hole solution given in Ref. \cite{WuWuWuYuPRD2021}. This rotating spacetime is described by the following spacetime metric in Boyer-Lindquist coordinates
	\begin{equation}
		ds^2 = - \frac{\Delta}{\varrho^2 } X^2+  \frac{\varrho ^2}{\Delta}dr^2 + \frac{\varrho ^2}{\Delta_\theta} d\theta^2 + \frac{\Delta_\theta \sin^2\theta}{\varrho ^2}Y^2,\label{eq:dKSAdSmetric} \
	\end{equation}
	where
	\begin{equation}
		X = dt - a \sin^2\theta \frac{d\phi}{\Xi}, ~~ Y = adt- (r^2-2dr-k^2+a^2 ) \frac{d\phi}{\Xi}, 
	\end{equation}
	\begin{equation}
		\Delta = (r^2-2dr -k^2+a^2)\left(1+\frac{r^2-2dr -k^2}{l^2} \right) -2m(r-d)+p^2+q^2, 
	\end{equation}
	\begin{equation}
		\Delta_\theta = 1-\frac{a^2}{l^2}\cos^2\theta, ~~\Xi =1-\frac{a^2}{l^2}, ~~
		\varrho^2 = r^2-2dr -k^2 + a^2\cos^2\theta.\label{eq:metricfunctiondKSAdS}\
	\end{equation}
	The  mass, spin, electric charge, magnetic (dyonic) charge, dilaton charge, and axion charge parameters 
   are respectively    $m, a, q, p, d, k$; $l$ is the radius of curvature of the AdS spacetime or the AdS length scale that is related to the cosmological constant, 
    \begin{equation}
    \Lambda= -\frac{3}{l^2}.\label{eq:linLambda}
    \end{equation}
    The dilaton and axion charges
	\begin{equation}
		d= \frac{p^2 -q^2}{2m}, ~~~~~ k =\frac{pq}{m} ,
	\end{equation}
     explicitly depend on the electromagnetic charges. 
	One can see that when one of the electromagnetic charges vanishes, the axion charge   also vanishes
    (yielding the non-dyonic solution~\cite{WuWuYuPRD2020} for  $p=0$).  
    When both electromagnetic charges possess the same value, the dilaton charge will vanish.  The electromagnetic field, its dual, dilaton scalar, and axion pseudoscalar fields of the dyonic Kerr-Sen-AdS black hole (\ref{eq:dKSAdSmetric}) are given by in the following forms:
	\begin{equation}
		A_\mu dx^\mu= \frac{q(r-p^2/m)}{\varrho^2}X-\frac{p\cos\theta}{\varrho^2}Y, ~~~
		B_\mu dx^\mu = \frac{p(r-p^2/m)}{\varrho^2}X+\frac{q\cos\theta}{\varrho^2}Y, \label{eq:BdKS}\
	\end{equation}
	\begin{equation}
		e^{\varphi} = \frac{r^2+(k+a\cos\theta)^2}{\varrho^2}, ~~~
		\chi = 2\frac{kr-d(k+a\cos\theta)}{r^2+(k+a\cos\theta)^2}.\ \label{eq:axiondKS}
	\end{equation}
	The dual of the gauge potential $A$ can be obtained from  $e^{-\varphi}\star F +\chi F = -dB$ where $B$ is the dual potential and $\star F$ is the Hodge dual of $F$ \cite{WuWuWuYuPRD2021}. 
	
	Since we want a static solution, we can set $a=0$, in which case the dyonic dilaton AdS black hole solution is given as follows
	\begin{equation}
		ds^2 = -\frac{\Delta}{\Sigma}dt^2 + \frac{\Sigma}{\Delta}dr^2 +\Sigma d\theta^2 +\Sigma \sin^2\theta d\phi^2,\label{eq:GMGHSAdSSol}
	\end{equation}
	where
	\begin{equation}
		\Delta = \left(\Sigma +\frac{\Sigma^2}{l^2}\right) -2m(r-d)+p^2+q^2, ~~~ \Sigma = r^2 - 2dr -k^2.
	\end{equation}
	The electromagnetic potential, its dual, dilaton scalar, and axion pseudoscalar fields then become
	\begin{equation}
		A_\mu dx^\mu = \frac{q(r-p^2/m)}{\Sigma}dt+p\cos\theta d\phi, ~~~
		B_\mu dx^\mu = \frac{p(r-p^2/m)}{\Sigma}dt-q\cos\theta d\phi, \label{eq:ElectroPot}\
	\end{equation}
	\begin{equation}
		e^{\varphi} = \frac{r^2+k^2}{\Sigma}, ~~~
		\chi = \frac{2k(r-d)}{r^2+k^2} , \label{eq:DilatonAxion}
	\end{equation}
and are obtained by setting $a=0$ in Eq. (\ref{eq:BdKS})-(\ref{eq:axiondKS}).

It is straightforward to show that if $a\neq 0$ then the spacetime has two horizons. However, for $a=0$ the metric function $\Delta$ in \eqref{eq:GMGHSAdSSol} has a root at $r_0=p^2/m$ and another root at a larger value of $r$.  However,
$\Sigma (r_0)=0$, and the spacetime is singular there; furthermore, the gauge fields and dilaton also diverge at $r=r_0$.
Consequently, this dyonic black hole has only a single horizon at $r > r_0$ and no extremal counterpart.
	
Several other interesting solutions emerge as special cases. In addition to the   non-dyonic $p=0$ recover the  Schwarzschild-AdS black hole for  $q=p=0$. We can also obtain the dyonic dilaton dS black hole solution by analytically continuing $l^2 \rightarrow -l^2$. Note that we cannot reproduce the  Reissner-Nordstr{\"o}m-AdS solution from this solution since it comes from a different theory.
	
\section{Generalized Mass/Energy Formulas}
\label{sec:massform}
	
\subsection{Extended Bulk Thermodynamics}\label{subsec:extended}
	
Since the extended bulk thermodynamics of the dyonic Kerr-Sen-AdS black hole has been computed \cite{WuWuWuYuPRD2021},   we obtain the thermodynamic parameters for the non-rotating version by setting $a=0$. As we  are interested in the mixed and CFT thermodynamics,   we need to restore   Newton's gravitational constant in the thermodynamic equations \cite{CongKubiznakPRL2021}.  We will exhibit the thermodynamic quantities of dyonic dilaton AdS black hole in terms of event horizon $r_+$ which is the largest root of $\Delta(r)$. 
     
     The Hawking temperature, Bekenstein-Hawking entropy, physical mass, electric charge, magnetic charge, electric potential, magnetic potential are given explicitly as follows
	\begin{equation}
		T =\frac{\kappa}{2\pi}= \frac{(r_+ - d)(2\Sigma_+ +l^2)-ml^2}{2\pi \Sigma_+ l^2}, ~~~
		S=\frac{A}{4G}=\frac{\pi}{G}\Sigma_+, ~~~ M= \frac{m}{G} ,\label{eq:SBHKSAdS}\
	\end{equation}
	\begin{equation}
		Q= \frac{q}{\sqrt{G}}, ~~~ P= \frac{p}{\sqrt{G}}, ~~~	\Phi = \frac{q(r_+  -p^2/m)}{\sqrt{G}\Sigma_+}, ~~~	\Psi = \frac{p(r_+ -p^2/m)}{\sqrt{G}\Sigma_+},
        \label{eq:PsiKSAdS}
	\end{equation}
    where $\Sigma_+=\Sigma(r_+)$.
	In extended bulk thermodynamics, the cosmological constant is treated as a thermodynamic quantity, known as thermodynamic pressure. This extended bulk thermodynamics is also known as black hole chemistry \cite{KubiznakMannCJP2014,KubiznakMannCQG2017,Mann:2025xrb}. For the dyonic dilaton AdS black hole, the thermodynamic pressure and its conjugate (thermodynamic volume) are 
	\begin{equation}
		V= \frac{4\pi}{3}(r_+ -d) \Sigma_+, ~~~\mathcal{P}=\frac{3}{8\pi G l^2}. \label{eq:VdKSAdS}
	\end{equation}
	Those thermodynamic quantities satisfy the following Bekenstein-Smarr mass formula \cite{WuWuWuYuPRD2021}
	\begin{equation}
		M = 2T S+\Phi Q+ \Psi P-2V \mathcal{P}. \label{eq:Smarrformula}
	\end{equation}
and it is straightforward to show that the first law \cite{WuWuWuYuPRD2021}
	\begin{equation}
		dM = TdS +\Phi dQ +\Psi dP +Vd\mathcal{P} \label{eq:firstlaw}
	\end{equation}
is satisfied for this dyonic dilaton AdS black hole. 
	In addition, the Christodoulou-Ruffini-like squared-mass is given in Ref. \cite{WuWuWuYuPRD2021}. By restoring the gravitational constant $G$, we can obtain
	\begin{eqnarray}
		M^2 = \frac{16G^3 \mathcal{P}^2 S^3}{9\pi}+\frac{4}{3}G\mathcal{P}S(P^2+Q^2)+\frac{4}{3\pi}G\mathcal{P}S^2 +\frac{S}{4\pi G} + \frac{P^2 +Q^2}{2G}.\label{eq:squaredmass}
	\end{eqnarray}
 One can derive this square-mass equation from $\Delta=0$ and by using  Eqs. (\ref{eq:SBHKSAdS})-(\ref{eq:VdKSAdS}). By using the expressions of the entropy in Eq. (\ref{eq:SBHKSAdS}) and pressure in Eq. (\ref{eq:VdKSAdS}) in terms of $\Lambda$ (\ref{eq:linLambda}), we can also write the squared-mass in terms of the cosmological constant and black hole area as follows
	\begin{equation}
		M^2 = \frac{A^3 \Lambda ^2}{2304\pi^3 G^2} -\frac{A^2\Lambda}{96\pi^2 G^2}+\frac{A}{16\pi G^2}-\frac{A\Lambda (P^2 + Q^2)}{24\pi G}+\frac{P^2 +Q^2}{2G}.\label{eq:squaredmassALambda}
	\end{equation}
	
	In the next section, we will consider the mixed thermodynamics where the central charge from the boundary thermodynamics is included. In order to do so, we need to treat the  gravitational constant $G$ as a thermodynamic parameter \cite{CongKubiznakPRL2021,AhmedCongPRL2023}.
    The first law of thermodynamics (\ref{eq:firstlaw}) when the gravitational constant is varied, is given by
	\begin{equation}
		dM = \frac{\kappa}{8\pi G}dA + \Phi dQ +\Psi dP - \frac{V}{8\pi G} d\Lambda -\alpha\frac{dG}{G} ~~~\text{where}~~~\alpha = M -\frac{1}{2}\Phi Q -\frac{1}{2}\Psi P, \label{eq:firstlawvariedG}
	\end{equation}
	When $P=0$, this first law is explicitly similar to that of Reissner-Nordstr{\"o}m-AdS black hole in Einstein-Maxwell theory. Using (\ref{eq:linLambda}), $T$ in Eq. (\ref{eq:SBHKSAdS}), and $\mathcal{P}$ in Eq. (\ref{eq:VdKSAdS}), the Smarr formula 
    (\ref{eq:Smarrformula}), 
      can be explicitly written as
	\begin{equation}
		M = \frac{\kappa A}{4\pi G} + \Phi Q +\Psi P +\frac{V\Lambda}{4\pi G}. \label{eq:SmarrvariedG}
	\end{equation}
	From Eq. (\ref{eq:squaredmassALambda}), we can derive the thermodynamic quantities in terms of $A$ and $\Lambda$ as follows
	\begin{eqnarray}
		\frac{\kappa}{8\pi G}&=&\left(\frac{\partial M}{\partial A}\right)_{Q,P, \Lambda, G} = \frac{1}{2M}\left(\frac{A^2\Lambda^2}{768\pi^3G^2} -\frac{A\Lambda}{48\pi^2 G^2}+\frac{1}{16\pi G^2}-\frac{\Lambda(P^2+Q^2)}{24\pi G}\right), \\
		\Phi &=&\left(\frac{\partial M}{\partial Q}\right)_{A,P, \Lambda, G} = \frac{1}{2M}\left(\frac{Q}{G}-\frac{A\Lambda Q}{12\pi G}\right),\\
		\Psi &=&\left(\frac{\partial M}{\partial P}\right)_{A,Q, \Lambda, G} = \frac{1}{2M}\left(\frac{P}{G}-\frac{A\Lambda P}{12\pi G}\right),\\
		\frac{V}{8\pi G} &=&-\left(\frac{\partial M}{\partial \Lambda}\right)_{A,Q,P,G} = \frac{1}{2M}\left(-\frac{A^3\Lambda}{1152\pi^3 G^2}+\frac{A^2}{96\pi^2 G^2}+\frac{A(P^2+Q^2)}{24\pi G}\right),\\
		\frac{\alpha}{G}&=&-\left(\frac{\partial M}{\partial G}\right)_{A,Q,P, \Lambda} = \frac{1}{2M}\left(\frac{A^3\Lambda^2}{1152\pi^3 G^3}-\frac{A^2\Lambda}{48\pi^2 G^3}+\frac{A}{8\pi G^3} -\frac{A\Lambda(P^2+Q^2)}{24\pi G^2}+\frac{P^2+Q^2}{2G^2}\right).\nonumber\\
		\
	\end{eqnarray}
	
	\subsection{Mixed Thermodynamics}\label{subsec:mixed}
	In this section, we want to transfer the extended bulk thermodynamic first law (\ref{eq:firstlaw}) into the mixed thermodynamic first law where   the boundary central charge plays a role as a thermodynamic quantity. In order to do so, we employ the holographic   relation between the central charge $C$, the bulk AdS length $l$, and the  gravitational constant $G$, which is given by \cite{CongKubiznakPRL2021,AhmedCongPRL2023}
	\begin{equation}
		C =\frac{\Omega_2 l^2}{16\pi G}.\label{eq:centralcharge}
	\end{equation}
	where $\Omega_2$ is the volume of a 2-sphere, which is equal to $4\pi$ for this black hole. By using the relation (\ref{eq:VdKSAdS}), we can get the gravitational constant as a function of the boundary central charge and the bulk thermodynamic pressure as
	\begin{equation}
		G = \frac{1}{4}\sqrt{\frac{3}{2\pi C \mathcal{P}}}.\label{eq:GinCP}
	\end{equation}
	
	Now, by using (\ref{eq:GinCP}), we can write the squared-mass formula (\ref{eq:squaredmassALambda}) in this mixed thermodynamics as
	\begin{eqnarray}
		M^2 = 2\left(P^2+Q^2\right)\sqrt{\frac{2\pi C \mathcal{P}}{3}} + S \sqrt{\frac{2C\mathcal{P}}{3\pi}} + S\left(P^2+Q^2\right)\sqrt{\frac{\mathcal{P}}{6\pi C}} +\frac{S^2}{\pi}\sqrt{\frac{\mathcal{P}}{6\pi C}} + \frac{S^3}{8\pi^2 C}\sqrt{\frac{\mathcal{P}}{6\pi C}}.\ \label{eq:squaredmassmixed}
	\end{eqnarray}
Furthermore,  inserting \eqref{eq:GinCP} into \eqref{eq:firstlawvariedG} yields the following first law in mixed thermodynamics,
	\begin{equation}
		dM = TdS + \Phi dQ +\Psi dP +V_c d\mathcal{P} +\mu_c dC,\label{eq:firstlawmixed}
	\end{equation}
where from \eqref{eq:squaredmassmixed} the thermodynamic quantities are
	\begin{eqnarray}
		T &=& \left( \frac{\partial M}{\partial S}\right)_{Q,P,\mathcal{P},C} =\frac{1}{2M} \left(\sqrt{\frac{2C\mathcal{P}}{3\pi }} + (P^2+Q^2)\sqrt{\frac{\mathcal{P}}{6\pi C}} + \frac{S}{\pi}\sqrt{\frac{2\mathcal{P}}{3\pi C}} + \frac{S^2}{8\pi^2 C}\sqrt{\frac{3\mathcal{P}}{2\pi C}}\right), \\
		\Phi &=& \left(\frac{\partial M}{\partial Q} \right)_{S,P,\mathcal{P},C} =\frac{1}{2M}\left(4Q\sqrt{\frac{2\pi C \mathcal{P}}{3}} +SQ\sqrt{\frac{2\mathcal{P}}{3\pi C}} \right)  , \\
		\Psi &=& \left(\frac{\partial M}{\partial P} \right)_{S,Q,\mathcal{P},C} =\frac{1}{2M}\left(4P\sqrt{\frac{2\pi C \mathcal{P}}{3}} +SP\sqrt{\frac{2\mathcal{P}}{3\pi C}} \right), \\
		V_c &=& \left(\frac{\partial M}{\partial \mathcal{P}} \right)_{S,Q,P,C} =\frac{1}{2M}  \left((P^2+Q^2)\sqrt{\frac{2\pi C}{3\mathcal{P}}} + S\sqrt{\frac{C}{6\pi \mathcal{P}}}+\frac{S(P^2+Q^2)}{2}\sqrt{\frac{1}{6\pi C \mathcal{P}}} \right. \nonumber\\
		& + &\left. \frac{S^2}{2\pi}\sqrt{\frac{1}{6\pi C \mathcal{P}}} +\frac{S^3}{16\pi^2 C}\sqrt{\frac{1}{6\pi C \mathcal{P}}} \right),\\
		\mu_c &=& \left(\frac{\partial M}{\partial C} \right)_{S,Q,P,\mathcal{P}} =\frac{1}{2M} \left((P^2+Q^2)\sqrt{\frac{2\pi\mathcal{P}}{3C}} + S\sqrt{\frac{\mathcal{P}}{6\pi C}}-\frac{S(P^2+Q^2)}{2C}\sqrt{\frac{\mathcal{P}}{6\pi C}}\right. \nonumber\\
		& - & \left. \frac{S^2}{2\pi C}\sqrt{\frac{\mathcal{P}}{6\pi C}} - \frac{S^3}{16\pi^2 C^2}\sqrt{\frac{3\mathcal{P}}{2\pi C}}\ \right) .\
	\end{eqnarray}
	where $V_c$ and $\mu_c$ are the new thermodynamic volume and the chemical potential that can also be written as
	\begin{equation}
		V_c = \frac{M}{4\mathcal{P}}, ~~~ \mu_c = \frac{2\mathcal{P}\left(V_c - V\right)}{2C}= \frac{M}{4C}-\frac{V\mathcal{P}}{C}.\label{eq:newVandmu}
	\end{equation}
    
	Note that in order to find Eq. (\ref{eq:firstlawmixed}), we have applied the variation of the gravitational  constant with respect to the central charge and the pressure, $dG/G = -d\mathcal{P}/(2\mathcal{P})-dC/(2C)$. Fixing $G$ determines variations of $C$ in terms of those in $\mathcal{P}$, whereas fixing the central charge determines variations of $G$ in terms of those in $\mathcal{P}$. In this mixed thermodynamics, by using Eq. (\ref{eq:newVandmu}) we can find the following Smarr-like formula
	\begin{equation}
		\frac{3}{2}M = 2TS + \Phi Q +\Psi P +2\mu_c C. \label{eq:Smarmixed}
	\end{equation}

	\subsection{CFT Thermodynamics}\label{subsec:CFT}
	In this section, we will investigate the CFT thermodynamics that is dual to the extended bulk thermodynamics of the dyonic dilaton AdS black hole. We begin by obtaining the energy formula in this CFT thermodynamics. 
    
    The dual CFT lives on the conformal boundary of the asymptotically AdS spacetime where the metric is given by the boundary metric of the dual asymptotically AdS spacetime up to a Weyl rescaling. We set the boundary metric to be
	\begin{equation}
		ds^2 = \omega^2(-dt^2 +l^2d\Omega^2_2),\label{eq:CFTmetric}\
	\end{equation}
	where $\omega$ is  an arbitrary dimensionless conformal factor and $d\Omega^2_2$ is the line element of a 2-sphere. The conformal factor can be chosen to be $\omega=R/l$ where $R$ is the curvature radius of the boundary. This metric is just like the Einstein static Universe (up to a constant Weyl factor) which is the standard product metric on $\mathbb{R}\times S^2$ \cite{GibbonsPerryCQG2005,GibbonsPerryPRL2005}. The spatial volume of the boundary sphere is proportional to $(\omega l)^2$. Hence, we can choose  $\tilde{V}= \Omega_2(\omega l)^2$. This volume is the conjugate of the boundary pressure $\tilde{\mathcal{P}}$.
	
	To obtain the energy formula or the Euler equation in the boundary, we  apply the following AdS/CFT dictionary
\cite{AhmedCongPRL2023} 
	\begin{equation}
		\tilde{S}=S, ~~E = \frac{M}{\omega}, ~~ \tilde{T} =\frac{T}{\omega}, ~~\tilde{\Phi} =\frac{\Phi \sqrt{G}}{\omega l},~~ \tilde{\Psi}=\frac{\Psi \sqrt{G}}{\omega l}, ~~\tilde{Q}=\frac{Ql}{\sqrt{G}},~~\tilde{P}=\frac{Pl}{\sqrt{G}}.\label{eq:AdSCFTdictionary}
	\end{equation}
  extended for non-vanishing magnetic charge. It is worth noting that the electromagnetic charges and their conjugates are rescaled by the AdS radius. Rather than set $R/l=1$ \cite{CongKubiznakPRL2021,CongKubiznakJHEP2022}, we shall keep this ratio general; this does not affect the mass formula in the CFT. From the first law of bulk thermodynamics with variable $G$ (\ref{eq:firstlawvariedG}) and using \eqref{eq:AdSCFTdictionary}, we   obtain 
	\begin{equation}
		dE = \tilde{T}d\tilde{S}+\tilde{\Phi}d\tilde{Q}+\tilde{\Psi}d\tilde{P}+\mu dC - \tilde{\mathcal{P}}d\tilde{V},\label{eq:CFTfirstlaw}\
	\end{equation}
    for the first law of CFT thermodynamics. The equation of state is
    \begin{equation}
    E=2\tilde{\mathcal{P}}\tilde{V}, \label{eq:EoSCFT}    
    \end{equation} 
    and the following Euler equation \cite{CongKubiznakPRL2021,CongKubiznakJHEP2022}
	\begin{equation}
		E = \tilde{T}\tilde{S}+\tilde{\Phi}\tilde{Q}+\tilde{\Psi}\tilde{P}+\mu C,\label{eq:EulerEq}
	\end{equation}
holds  in the large-$N$ gauge theory; see Appendix \ref{sec:appA} for the derivation. This energy formula differs from the standard one in extended bulk thermodynamics which does not contain a work term, $\tilde{\mathcal{P}}\tilde{V}$.
	
We can also construct the squared-energy formula in CFT thermodynamics. Employing the AdS/CFT dictionary (\ref{eq:AdSCFTdictionary}) and the CFT equation of state, we find
	\begin{equation}
		E^2 =\frac{(4\pi C + \tilde{S})\left\{\tilde{S}^2+2\pi^2 (\tilde{Q}^2+\tilde{P}^2) +4\pi C \tilde{S}\right\}}{4\pi^2 C  \tilde{V}},\label{eq:squaredE}\
	\end{equation}
 for the squared-energy formula, 
	where from (\ref{eq:centralcharge}) we have used $\tilde{\mathcal{P}}=\frac{24\pi \omega^4 C}{\tilde{V^2}}$ and $G = \frac{\tilde{V}}{16\pi \omega^2 C}$. From \eqref{eq:squaredE}  we obtain 
	\begin{eqnarray}
		\tilde{T} &=& \left( \frac{\partial E}{\partial \tilde{S}}\right)_{\tilde{Q},\tilde{P},\tilde{V},C} =\frac{1}{2E} \left(\frac{3\tilde{S}^2 + 16\pi C \tilde{S} + 2\pi^2  (\tilde{Q}^2 +\tilde{P}^2) + 16\pi^2 C^2 }{4\pi^2 C  \tilde{V}} \right),\label{eq:TemperatureCFT}\\
		\tilde{\Phi} &=& \left(\frac{\partial E}{\partial \tilde{Q}} \right)_{\tilde{S},\tilde{P},\tilde{V},C} =\frac{1}{2E}\left(\frac{\tilde{Q}(4\pi C +\tilde{S})}{C\tilde{V}} \right)  , \label{eq:PhiCFT}\\
		\tilde{\Psi} &=& \left(\frac{\partial E}{\partial \tilde{P}} \right)_{\tilde{S},\tilde{Q},\tilde{V},C} =\frac{1}{2E}\left(\frac{\tilde{P}(4\pi C +\tilde{S})}{C\tilde{V}} \right)  , \label{eq:PsiCFT}\\
		\mu &=& \left(\frac{\partial E}{\partial C} \right)_{\tilde{S},\tilde{Q},\tilde{P},\tilde{V}} =\frac{1}{2E} \left(\frac{16\pi^2 C^2 \tilde{S}  - \tilde{S}^3 -2\pi^2 \tilde{S}(\tilde{Q}^2+\tilde{P}^2)}{4\pi^2 C^2  \tilde{V}}\right), \label{eq:muCFT}\\
		\tilde{\mathcal{P}} &=& -\left(\frac{\partial E}{\partial \tilde{V}} \right)_{\tilde{S},\tilde{Q},\tilde{P},C} = \frac{1}{2E} \left(\frac{(4\pi C + \tilde{S})\left(\tilde{S}^2 +2\pi^2 (\tilde{Q}^2 + \tilde{P}^2) + 4\pi C \tilde{S}\right)}{4\pi^2 C \tilde{V}^2} \right). \label{eq:pCFT}\
	\end{eqnarray}
for the various thermodynamic quantities.  

Besides the pair ($\tilde{T}, \tilde{S}$) there are four pairs of conjugate thermodynamic variables in the CFT description of dyonic dilaton AdS black holes, which are ($\tilde{\Phi}, \tilde{Q}$), ($\tilde{\Psi}, \tilde{P}$), ($\tilde{\mathcal{P}}, \tilde{V}$)and ($\mu, C$). This yields a total of sixteen possible emsembles. We shall first examine the fixed $\tilde{V}$ ensembles and then the 
fixed $\tilde{\mathcal{P}}$ ensembles. 
    
\section{Thermodynamic Ensembles in the CFT for constant $\tilde{V}$}\label{sec:ensemble_constantV}

    There are eight distinct 
    fixed $\tilde{V}$ ensembles
	   whose  
    free energies are
	\begin{eqnarray}
		&&\text{fixed}~(\tilde{Q},\tilde{P},\tilde{V},C): ~~~ F_1 = E - \tilde{T}\tilde{S}=\tilde{\Phi}\tilde{Q}+\tilde{\Psi}\tilde{P}+\mu C, \label{eq:F1}\\
		&&\text{fixed}~(\tilde{Q},\tilde{P},\tilde{V},\mu): ~~~F_{2} = E - \tilde{T}\tilde{S} -\mu C=\tilde{\Phi}\tilde{Q}+\tilde{\Psi}\tilde{P}, \label{eq:F2}\\
		&&\text{fixed}~(\tilde{\Phi},\tilde{P},\tilde{V},C): ~~~F_{3} = E - \tilde{T}\tilde{S}-\tilde{\Phi}\tilde{Q}=\tilde{\Psi}\tilde{P}+\mu C, \label{eq:F3}\\
		&&\text{fixed}~(\tilde{\Phi},\tilde{P},\tilde{V},\mu): ~~~ F_{4} = E - \tilde{T}\tilde{S}-\tilde{\Phi}\tilde{Q}-\mu C=\tilde{\Psi}\tilde{P}, \label{eq:F4}\\
		&&\text{fixed}~(\tilde{\Phi},\tilde{\Psi},\tilde{V},\mu): ~~~ F_{5} = E - \tilde{T}\tilde{S}-\tilde{\Phi}\tilde{Q} -\tilde{\Psi}\tilde{P}-\mu C= 0, \label{eq:F5}\\
		&&\text{fixed}~(\tilde{Q},\tilde{\Psi},\tilde{V},\mu): ~~~ F_{6}= E -\tilde{T}\tilde{S}-\tilde{\Psi}\tilde{P}-\mu C=\tilde{\Phi}\tilde{Q}, \label{eq:F6}\\
		&&\text{fixed}~(\tilde{Q},\tilde{\Psi},\tilde{V},C): ~~~ F_{7} = E - \tilde{T}\tilde{S}-\tilde{\Psi}\tilde{P}=\tilde{\Phi}\tilde{Q}+\mu C, \label{eq:F7}\\
		&&\text{fixed}~(\tilde{\Phi},\tilde{\Psi},\tilde{V},C): ~~~ F_{8} = E - \tilde{T}\tilde{S}-\tilde{\Phi}\tilde{Q}-\tilde{\Psi}\tilde{P}=\mu C, \label{eq:F8}
		 \label{eq:F16}\
	\end{eqnarray}
one of which ($F_5$) is vanishing and so does not warrant further consideration.  Furthermore, $F_6$ and $F_7$ are qualitatively similar to $F_4$ and $F_3$ respectively due to the symmetry between $\tilde{P}$ and $\tilde{Q}$ in equations \eqref{eq:squaredE} -- \eqref{eq:pCFT}, and so they likewise do not warrant further analysis. 

For all ensembles, the temperature will be given by
\eqref{eq:TemperatureCFT}, which can be written as
\begin{equation}
		\tilde{T} = \frac{3\tilde{S}^2 + 16\pi C \tilde{S} + 2\pi^2 (\tilde{Q}^2 +\tilde{P}^2) + 16\pi^2 C^2}{4\pi\sqrt{C\tilde{V}\left[(4\pi C+\tilde{S})\left(\tilde{S}^2+4\pi C\tilde{S}+2\pi^2 (\tilde{P}^2+\tilde{Q}^2)\right)\right]}}, \label{eq:Texplicit}
	\end{equation}
using \eqref{eq:squaredE}, though in some cases some of the variables $(\tilde{Q},\tilde{P},C)$ will be understood as functions of $(\tilde{\Phi},\tilde{\Psi},\mu)$ depending on the choice of ensemble.  
 \begin{figure*}
		\centering
		\begin{tabular}{c c c}
			\includegraphics[width=.35\linewidth,valign=m,margin=-0.4cm -0.1cm]{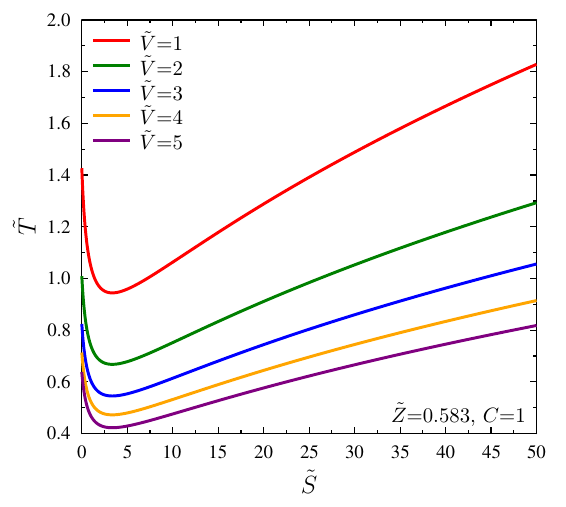} & \includegraphics[width=.35\linewidth,valign=m,margin=-0.35cm -0.1cm]{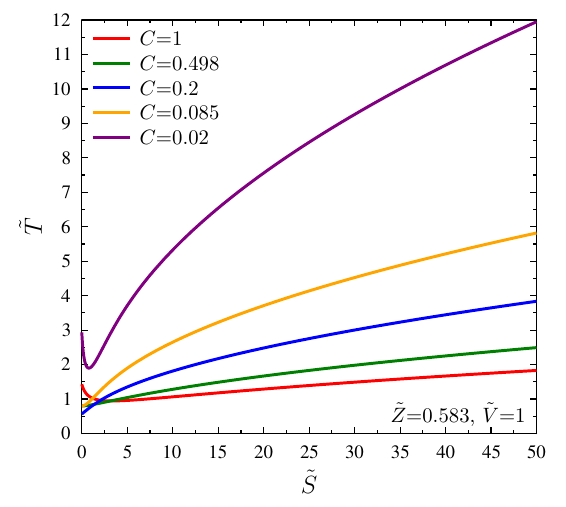} & \includegraphics[width=.35\linewidth,valign=m,margin=-0.3cm -0.1cm]{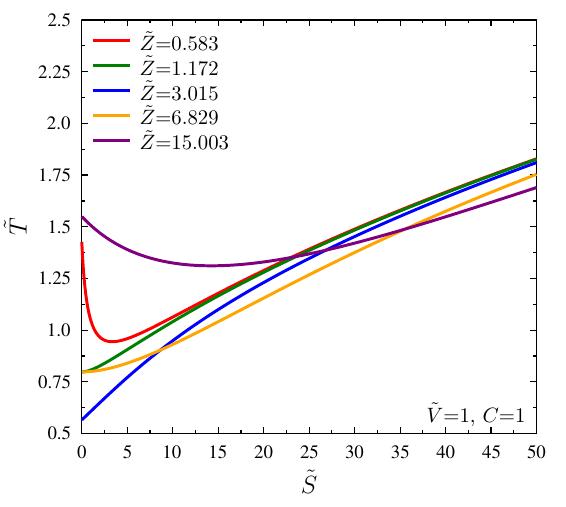}\\
			(a) & (b) & C\
		\end{tabular}
		\caption{Plots of temperature $\tilde{T}$ vs entropy $\tilde{S}$ for the fixed $(\tilde{Q},\tilde{P},\tilde{V},C)$ ensemble. (a) We fix $Z=0.583,C=1$ for various choices of $\tilde{V}=1,2,3,4,5$. (b) We fix $Z=0.583, \tilde{V}=1$ for various choices of $C=1, 0.498, 0.2, 0.085, 0.02$. (c) We fix $\tilde{V},C=1$ for various choices of $Z=  0.583, 1.172, 3.015, 6.829, 15.003$.}
		\label{fig:temperature}
	\end{figure*}  
There will be a minimum value, $T_{min}$, of the temperature 
	\begin{equation}
		\left(\frac{\partial \tilde{T}}{\partial \tilde{S}}\right)_{\tilde{V},\tilde{Q},\tilde{P},C}=0,\label{eq:dT=0}\
	\end{equation}
provided the ratio $z=\tilde{Z}/C$
is not in the range
\begin{equation}
    2\left(2-\sqrt{2}\right) \leq z \leq 2\left(2+\sqrt{2}\right) \; \label{eq:boundofTmin},
\end{equation}   
where $\tilde{Z}=\sqrt{\tilde{Q}^2+\tilde{P}^2}$. For values of $z$ in this range, the temperature monotonically increases from a finite value at $\tilde{S}=0$.

Plots of the temperature \eqref{eq:Texplicit} as a function of the entropy in CFT thermodynamics for a range of values of the other variables are shown in Figs. \ref{fig:temperature} (a) - (c). 
We see that when a minimum is present, there are two distinct branches of states. One is a low-entropy branch dual to small black hole solutions, and the other is a   high-entropy branch  dual to  large black hole solutions.  These two branches occur provided the minimum temperature is at $\tilde{S} > 0$; otherwise   there exists only one branch of states. This behaviour is fully analogous to that of the Schwarzschild AdS black hole, which likewise has a single minimal value of the temperature as a function of the entropy. Fig. \ref{fig:temperature} (a) illustrates a choice of $\tilde{Z},C$ for which  both branches exist. The minimum temperature decreases with increasing $\tilde{V}$.  The other panels show scenarios at fixed $\tilde{V}$ where the minimum temperature occurs at $\tilde{S}=0$ for particular values of $\tilde{Z},C$. In Fig. \ref{fig:temperature} (b), we set For $\tilde{Z}=0.583,\tilde{V}=1$, so the single branch occurs when $0.498\leq C\leq 0.085$ (green, blue, orange lines). In Fig. \ref{fig:temperature} (c), we set For $C=1,\tilde{V}=1$, so the single branch occurs when $1.172\leq \tilde{Z}\leq 6.828$ (green, blue, orange lines).
    
\subsection{Ensemble with fixed $(\tilde{Q},\tilde{P},\tilde{V},C)$}
\label{subsec:QPVC}
    
We begin by considering the ensemble with fixed electric charge $\tilde{Q}$,   magnetic charge $\tilde{P}$,   spatial volume $\tilde{V}$, and   central charge $C$. The thermodynamic potential in this ensemble is the Helmholtz free energy (\ref{eq:F1}), $F_1 = E- \tilde{T} \tilde{S}$ by the CFT first law. The differential of $F_1$ satisfies
	\begin{equation}
		dF_1 = dE- \tilde{T}d\tilde{S} -\tilde{S}d\tilde{T}=-\tilde{S}d\tilde{T}+\tilde{\Phi}d\tilde{Q}+\tilde{\Psi}d\tilde{P}+\mu dC - \tilde{\mathcal{P}}d\tilde{V}, \label{eq:dF1}\
	\end{equation}
implying that at fixed $(\tilde{Q},\tilde{P},\tilde{V},C)$, the free energy $F_1$ is stationary. The free energy $F_1$  is explicitly given by
	\begin{equation}
		F_1 = \frac{2\pi^2 \tilde{S}(\tilde{Q}^2+\tilde{P}^2) -\tilde{S}^3 +16\pi^2 C^2 \tilde{S} +16\pi^3 C (\tilde{Q}^2+\tilde{P}^2) }{4\pi\sqrt{C\tilde{V}\left[(4\pi C+\tilde{S})\left(\tilde{S}^2+4\pi C\tilde{S}+2\pi^2 (\tilde{P}^2+\tilde{Q}^2)\right)\right]}}.\label{eq:F1explicit}
	\end{equation}

	\begin{figure*}
		\centering
		\begin{tabular}{c c c}
			\includegraphics[width=.35\linewidth,valign=m,margin=-0.4cm -0.1cm]{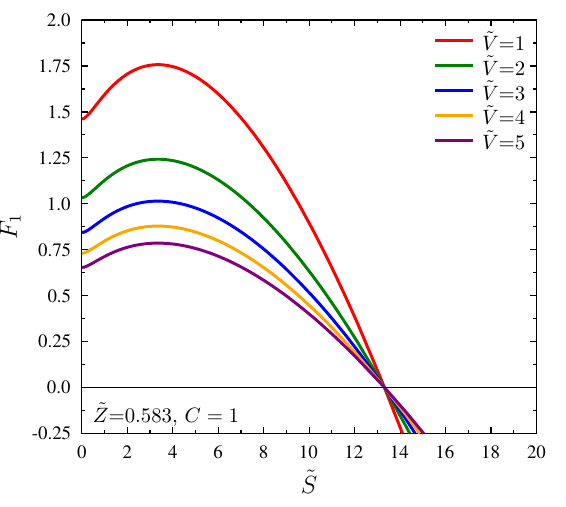} & \includegraphics[width=.35\linewidth,valign=m,margin=-0.35cm -0.1cm]{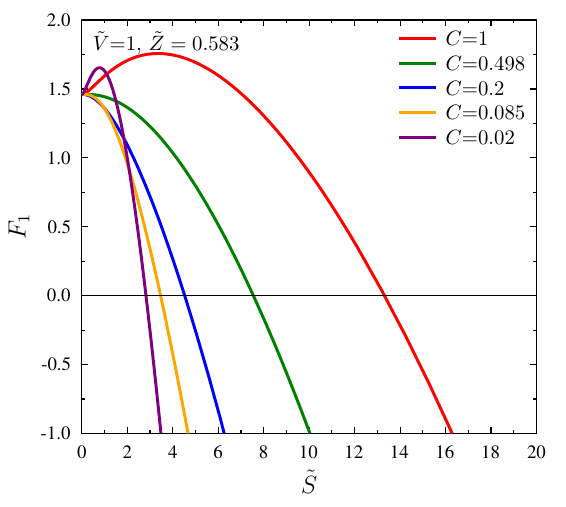} &
			\includegraphics[width=.35\linewidth,valign=m,margin=-0.35cm -0.1cm]{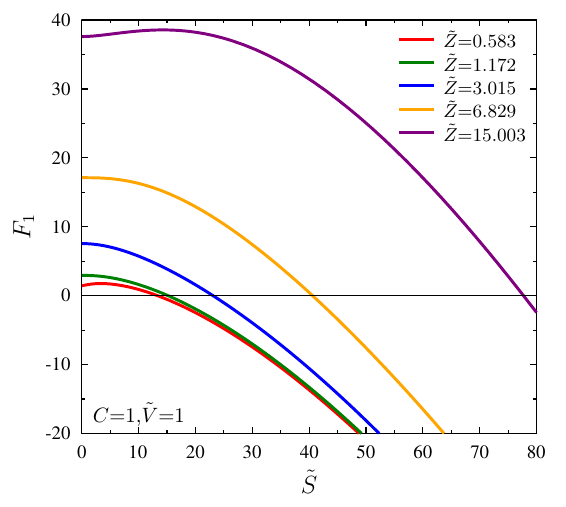} \\
			(a) & (b) & (c) \\
            \includegraphics[width=.35\linewidth,valign=m,margin=-0.4cm -0.1cm]{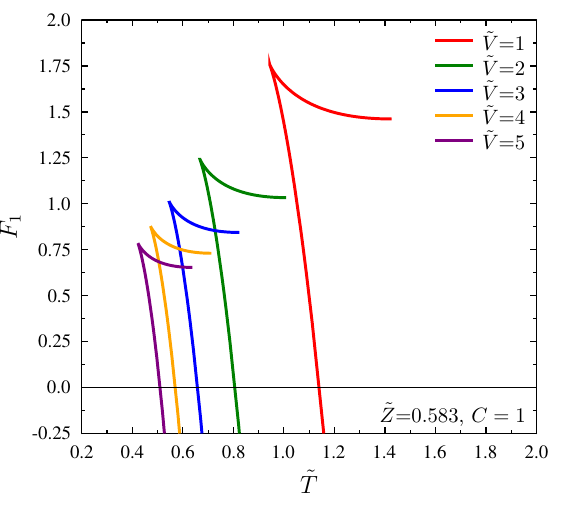} & \includegraphics[width=.35\linewidth,valign=m,margin=-0.35cm -0.1cm]{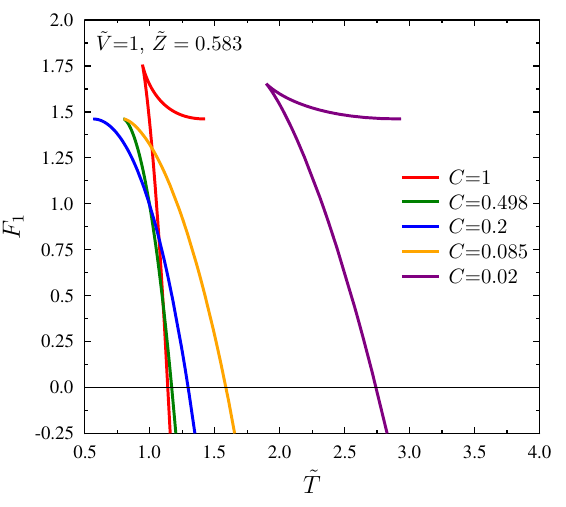} &
			\includegraphics[width=.35\linewidth,valign=m,margin=-0.35cm -0.1cm]{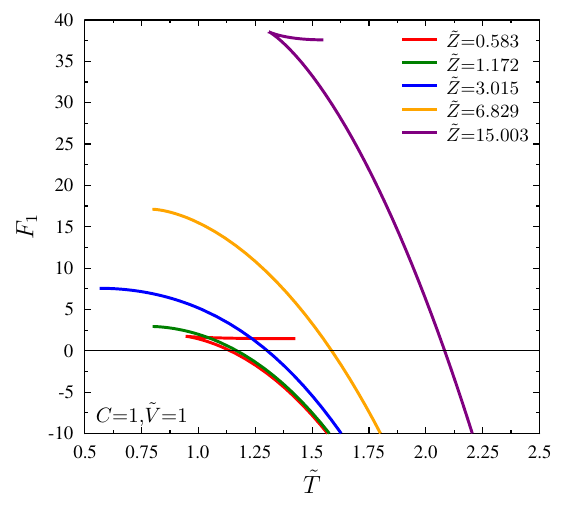} \\
			(d) & (e) & (f) \
		\end{tabular}
		\caption{\textit{Top}: Plots of  free energy $F_1$ vs entropy $\tilde{S}$ for the fixed  $(\tilde{Q},\tilde{P},\tilde{V},C)$ ensemble for various choices of (a) the spatial volume  $\tilde{V}=1,2,3,4,5$, (b) central charge $C=1, 0.498, 0.2, 0.085, 0.02$, and (c) $\tilde{Z}=0.583, 1.1715, 3.015, 6.829, 15.003$.    
        \textit{Bottom}: Plots of  free energy $F_1$ vs temperature $\tilde{T}$ with the similar choices of parameters as top panels.}
		\label{fig:F1vsST}
	\end{figure*}
In Figs. \ref{fig:F1vsST} (a) - (c), we  plot   $F_1$ as a function of $\tilde{S}$ and as a function of (parametrically) $\tilde{T}$ in Figs. \ref{fig:F1vsST} (d) - (f) for various choice $\tilde{V},C,\tilde{Z}$. The plots as a function of temperature yield curves with cusps that are reminiscent of those for Schwarzschild-AdS black holes.  However, there is no Hawking-Page phase transition to radiation when $F_1=0$ since the charges are fixed and conserved. Furthermore, the endpoints of the upper parts of the curves in Fig.~\ref{fig:F1vsST} correspond to the zero-entropy state as $r\to p^2/m$ -- this is the smallest possible black hole.
    
We  see that the free energy exhibits a bifurcation point from which two branches or states emerge. This bifurcation point or cusp denotes the minimum temperature $T_{min}$. 
   Beginning at $T_{min}$, as the temperature increases  the free energy of both branches  decreases, with the stable branch of higher entropy having lower free energy.  For temperatures larger than
\begin{equation}
		\hat{T} = \frac{ 2\pi^2 (\tilde{Q}^2 +\tilde{P}^2) + 16\pi^2 C^2}{4\pi\sqrt{8\pi^3 C^2\tilde{V} \left( \tilde{P}^2+\tilde{Q}^2\right) }} , \label{eq:Texplicit}
	\end{equation}
 only the single stable branch exists.
   
All such plots in Fig. \ref{fig:F1vsST} (d) - (f) have the familiar cusp behaviour, though for some values of the parameters the upper branch does not appear.  This single-valued free energy occurs for certain (bounded) ranges of the central charge $C$ and $\tilde{Z}$ provided by the Eq. (\ref{eq:boundofTmin}). In this single-valued free energy, only one entropy state exists which is globally stable.  In Figs. \ref{fig:F1vsST} (b) and (e) for fixed value of $\tilde{V},\tilde{Z}$, the single-valued $F_1$ occurs when $0.085\leq C \leq 0.498$. In Fig. \ref{fig:F1vsST} (c) and (f) for fixed value of $\tilde{V}, C$, the single-valued $F_1$ occurs when $1.172\leq \tilde{Z}\leq 6.829$. The change from two-branches to single-valued free energy occurs also in the CFT thermodynamics of charged AdS black holes in the ensemble with fixed electric potential, spatial volume, and central charge \cite{CongKubiznakJHEP2022}.

	\subsection{Ensemble with fixed ($\tilde{Q},\tilde{P},\tilde{V},\mu$)}
	\label{subsec:QPVmu}

    
	\begin{figure*}
		\centering
		\begin{tabular}{c c c}
			\includegraphics[width=.35\linewidth,valign=m,margin=-0.4cm -0.1cm]{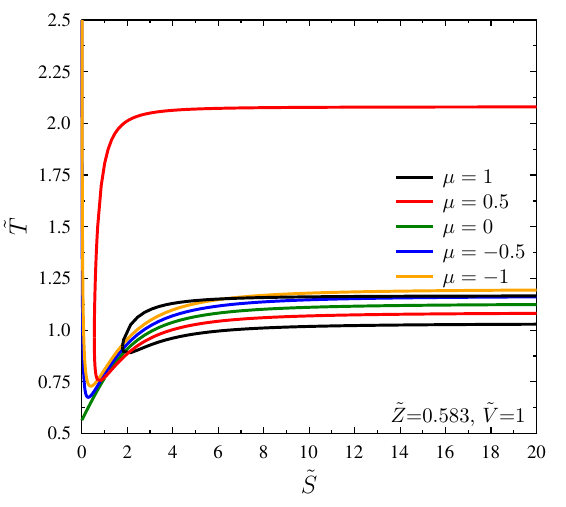} & \includegraphics[width=.35\linewidth,valign=m,margin=-0.3cm -0.1cm]{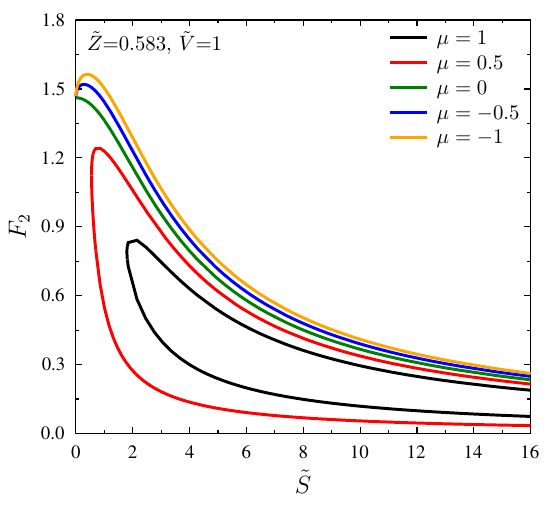} & \includegraphics[width=.35\linewidth,valign=m,margin=-0.3cm -0.1cm]{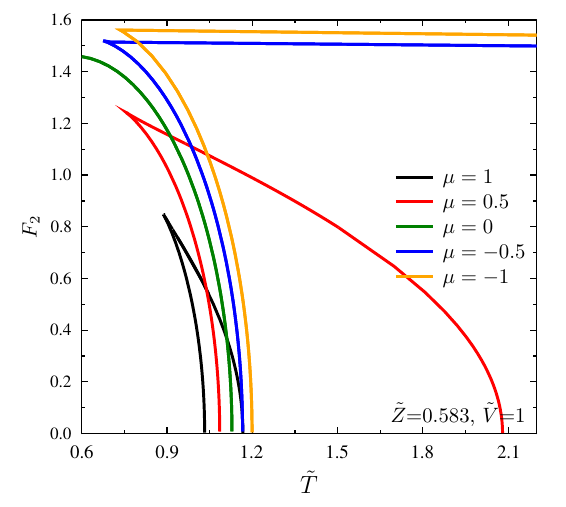} \\
			(a) & (b) & (c)\
		\end{tabular}
		\caption{Plot of (a) temperature $\tilde{T}$ vs entropy $\tilde{S}$, (b) free energy $F_2$ vs entropy $\tilde{S}$, and (c) free energy $F_2$ vs temperature $\tilde{T}$ with the variation of $\mu$ as $-1, -0.5, 0, 0.5, 1$ for ensemble at fixed ($\tilde{Q},\tilde{P},\tilde{V},\mu$). We set the parameters $Z=0.583$ and $\tilde{V}=1$.}
		\label{fig:F2vsST}
	\end{figure*}
	We next carry out the study of the ensemble with fixed electric charge $\tilde{Q}$, the magnetic charge $\tilde{P}$, the spatial volume $\tilde{V}$, and the chemical potential $\mu$. The free energy in this ensemble is given by Eq. (\ref{eq:F2}), $F_2 = E- \tilde{T} \tilde{S} -\mu C$. The differential of $F_2$ satisfies the following relation
	\begin{equation}
		dF_2 = dE- \tilde{T}d\tilde{S} -\tilde{S}d\tilde{T} - \mu dC -Cd\mu=-\tilde{S}d\tilde{T}-Cd\mu+\tilde{\Phi}d\tilde{Q}+\tilde{\Psi}d\tilde{P} - \tilde{\mathcal{P}}d\tilde{V}. \label{eq:dF2}\
	\end{equation}
	At fixed $(\tilde{Q},\tilde{P},\tilde{\mathcal{V}},\mu)$, the free energy $F_2$ is stationary as we can observe from above equation. The explicit form of $F_2$ is given by
	\begin{equation}
		F_2 = \frac{\pi(\tilde{P}^2+\tilde{Q}^2)(4\pi C+\tilde{S})}	{\sqrt{C\tilde{V}\left[(4\pi C+\tilde{S})\left(\tilde{S}^2+4\pi C\tilde{S}+2\pi^2 (\tilde{P}^2+\tilde{Q}^2)\right)\right]}}, \label{eq:F2explicit}
	\end{equation}
 and is non-negative. We cannot show the explicit form of $F_2$ in terms of $\mu$ since we cannot obtain the analytic form of $C$ in terms of $\mu$. Nevertheless, we can show the plot of the free energy as given in Fig. \ref{fig:F2vsST} by solving Eq. (\ref{eq:muCFT}), so we have $C(\mu)$. From Eq. (\ref{eq:F2explicit}), we can see that $F_2$ is always positive. Hence, the first-order phase transition cannot occur.
	
In Fig. \ref{fig:F2vsST}, we plot the temperature and free energy $F_2$ as functions of the entropy $\tilde{S}$ and $F_2$ as a function of $\tilde{T}$ for fixed  $\tilde{Z}=0.583$, and $\tilde{V}=1$. We see from Fig. \ref{fig:F2vsST} (a) that the temperature possesses a minimum value $T_{min}$ and is double-valued for $\mu>0$, whereas for $\mu<0$ it is single-valued. In this latter case there is a state either of high entropy (near the horizontal asymptote) or of low entropy (near the vertical asymptote).


Plots of the free energy $F_2$ as a function of $\tilde{S}$ and $\tilde{T}$ (parametrically) are depicted in Figs. \ref{fig:F2vsST} (b) and (c), respectively. From the latter diagram we see that only when $\mu=0$ is there a single phase; all other values of $\mu$ admit two different phases. The $\mu<0$ behaviour, shown in the blue and orange curves, is similar to the previous fixed $(\tilde{Q},\tilde{P},\tilde{V},C)$ ensemble, with a stable high-entropy branch, an unstable low-entropy branch, and a minimum temperature $T_{min}$ below which no solutions exist. Unlike the previous case, however, the free energy curve has a state that terminates at  $T_1$ where $F_2=0$. For temperatures larger than $T_1$, only the upper low-entropy branch exists; as the temperature decreases from $T>T_1$ to $T<T_1$, there will be a zeroth-order phase transition from the low-entropy upper branch to the high-entropy lower branch. To find $T_1$ and $T_2$, we  solve  Eq. (\ref{eq:muCFT}) numerically for  $C(\mu)$ and use it as input to the temperature (\ref{eq:Texplicit}). For $\mu=1$, we have $T_1 = 1.032$ and $T_2= 1.167$, whereas  for $\mu=0.5$, we have $T_1 = 1.085$ and $T_2=2.08$.

	\subsection{Ensemble with fixed $(\tilde{\Phi},\tilde{P},\tilde{V},C)$}
	\label{subsec:PhiPVC}
	
	\begin{figure*}
		\centering
		\begin{tabular}{c c c}
			\includegraphics[width=.35\linewidth,valign=m,margin=-0.4cm -0.1cm]{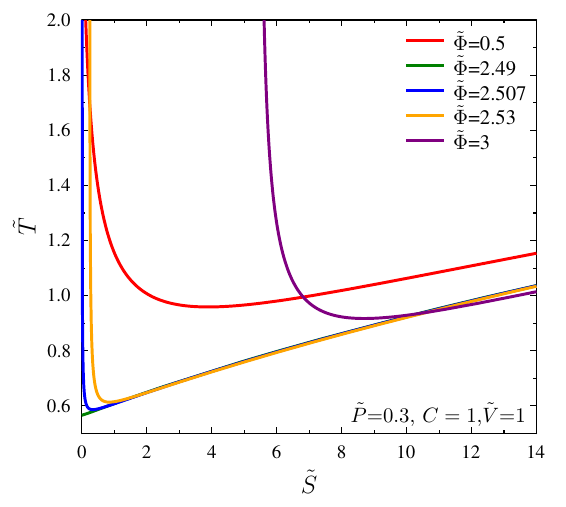} & \includegraphics[width=.35\linewidth,valign=m,margin=-0.4cm -0.1cm]{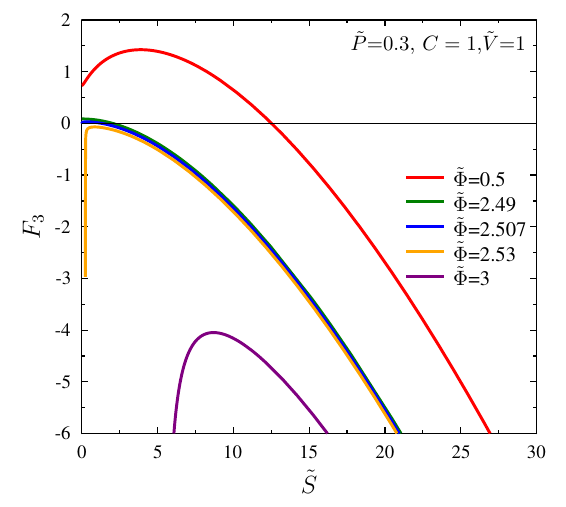} & \includegraphics[width=.35\linewidth,valign=m,margin=-0.3cm -0.1cm]{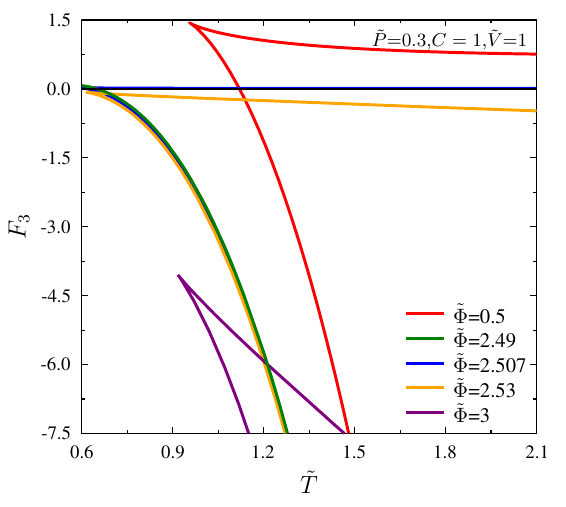} \\
			(a) & (b) & (c) \
		\end{tabular}
		\caption{Plot of (a) temperature $\tilde{T}$ vs entropy $\tilde{S}$, (b) free energy $F_3$ vs entropy $\tilde{S}$, and (c) free energy $F_3$ vs temperature $\tilde{T}$ for variation of $\tilde{\Phi}$ as $0.5, 2.49, 2.507, 2.53, 3$ for ensemble at fixed $(\tilde{\Phi},\tilde{P},\tilde{V},C)$. We set the parameters $\tilde{P}=0.3$, $C=1$, and $\tilde{V}=1$. }
		\label{fig:F3vsT}
	\end{figure*}
	
The next ensemble we consider is with fixed electric potential $\tilde{\Phi}$, magnetic charge $\tilde{P}$, spatial volume $\tilde{V}$, and central charge $C$. The free energy of this ensemble is $F_3 = E- \tilde{T}\tilde{S}-\tilde{\Phi}\tilde{Q}$ as given in Eq. (\ref{eq:F3}). This ensemble is qualitatively similar to the ensemble with fixed $(\tilde{Q},\tilde{\Psi},\tilde{V},C)$ with free energy $F_7$ (\ref{eq:F7}). The differential of $F_3$ is given by
	\begin{equation}
		dF_3 = dE- \tilde{T}d\tilde{S} -\tilde{S}d\tilde{T} -\tilde{\Phi}d\tilde{Q} -\tilde{Q}d\tilde{\Phi}=-\tilde{S}d\tilde{T}-\tilde{Q}d\tilde{\Phi}+\tilde{\Psi}d\tilde{P} +\mu dC - \tilde{\mathcal{P}}d\tilde{V} ,
        \label{eq:dF3}\
	\end{equation}
and so  $F_3$ is stationary at fixed $(\tilde{\Phi},\tilde{P},\tilde{V},C)$. Explicitly we obtain
	\begin{equation}
		F_3 = \frac{2\pi^2 \tilde{S}(\tilde{P}^2-\tilde{Q}^2(\tilde{\Phi})) -\tilde{S}^3 +16\pi^2 C^2 \tilde{S} +16\pi^3 C \tilde{P}^2 }{4\pi\sqrt{C\tilde{V}\left[(4\pi C+\tilde{S})\left(\tilde{S}^2+4\pi C\tilde{S}+2\pi^2 (\tilde{P}^2+\tilde{Q}^2(\tilde{\Phi}))\right)\right]}},\label{eq:F3explicit}
	\end{equation}
where
\begin{equation}
    \tilde{Q}(\tilde{\Phi}) 
    = \pm\frac{\tilde{\Phi}}{\pi} \sqrt{C\tilde{V} \frac{\tilde{S}^2+4\pi C\tilde{S}+2\pi^2 \tilde{P}^2}{\tilde{S}+4\pi C -2C\tilde{\Phi}^2 \tilde{V}} },
\end{equation}
is obtained by solving  (\ref{eq:squaredE}) and (\ref{eq:PhiCFT}).

We exhibit a plot of the temperature versus entropy
for various fixed $\tilde{\Phi}$  in Fig. \ref{fig:F3vsT} (a). Again there is a minimum temperature all possible choices of  $\tilde{\Phi}$ that separates the high- and low-entropy states. However there is a small range of $\Phi$ over which this minimum is at negative values of $S$. In Fig. \ref{fig:F3vsT} (a), with $\tilde{P}=0.3$, $C=1$, and $\tilde{V}=1$, this range is $\Phi \in (2.488665,2.49497)$;
the green line illustrates a case within this range.
 
Plots of  $F_3$ vs entropy and temperature are  given in Figs. \ref{fig:F3vsT} (b) and (c) respectively for various values of $\Phi$.  Although the CFT can electrically discharge (since the electric potential is fixed), it cannot make a transition to thermal AdS at $F_3=0$ since the magnetic charge $\tilde{P}$ is fixed and conserved. For temperatures below the value where the cusp occurs there are no solutions to the system. There is no cusp over the narrow range of $\Phi$ at which the temperature has no minimum for $\tilde{S} > 0$.

	\subsection{Ensemble with fixed $(\tilde{\Phi},\tilde{P},\tilde{V},\mu)$}
	\label{subsec:PhiPVmu}
	
	\begin{figure*}
		\centering
		\begin{tabular}{c c c}
			\includegraphics[width=.34\linewidth,valign=m,margin=-0.4cm -0.1cm]{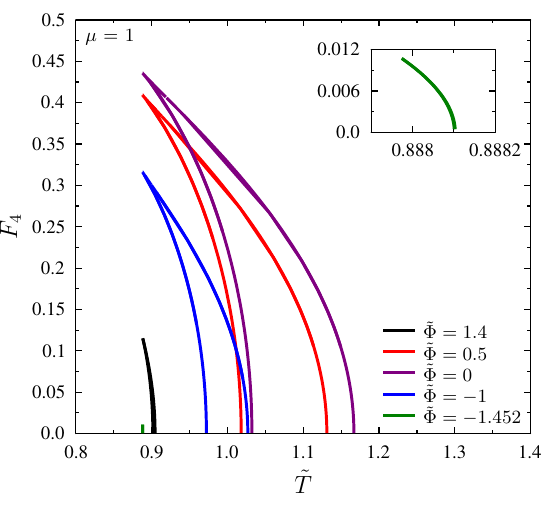} & \includegraphics[width=.34\linewidth,valign=m,margin=-0.2cm -0.1cm]{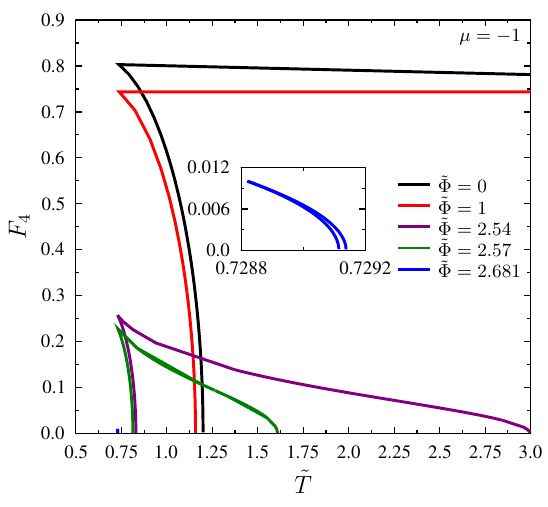} & \includegraphics[width=.34\linewidth,valign=m,margin=-0.3cm -0.1cm]{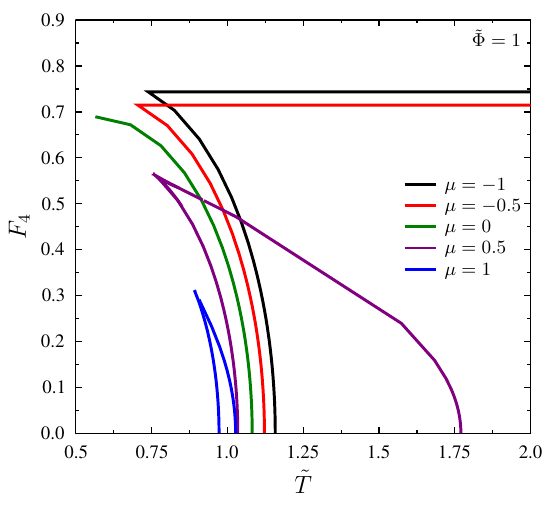}\\
			(a) & (b) & (c) \
		\end{tabular}
		\caption{Plot of free energy $F_4$ vs temperature with (a) $\mu=1$ with varied $\tilde{\Phi}$ as $-1.452, -1, 0, 0.5, 1.4$, (b) $\mu=-1$ with varied $\tilde{\Phi}$ as $0, 1, 2.54, 2. 57, 2.681$ and (c) $\tilde{\Phi}=1$ with varied $\mu$ as $-1, -0.5, 0, 0.5, 1$ for ensemble at fixed $(\tilde{\Phi},\tilde{P},\tilde{V},\mu)$. We set the parameters $\tilde{P}=0.3$ and $\tilde{V}=1$.
        The inset in panel (a) shows the case $\tilde{\Phi}=-1.452$, the minimal value of $\tilde{\Phi}$.
        }
		\label{fig:F4vsT}
	\end{figure*}
	The next ensemble will be the ensemble with fixed electric charge $\tilde{\Phi}$, the magnetic charge $\tilde{P}$, the spatial volume $\tilde{V}$, and the chemical potential $\mu$. In this ensemble, the free energy is given by Eq. (\ref{eq:F4}), $F_4 = E- \tilde{T} \tilde{S} -\tilde{\Phi}\tilde{Q} -\mu C$. This ensemble is qualitatively similar to the ensemble with fixed $(\tilde{Q},\tilde{\Psi},\tilde{V},\mu)$ with free energy $F_6$ (\ref{eq:F6}). We can find the differential of $F_4$, which satisfies
	\begin{equation}
		dF_4 = dE- \tilde{T}d\tilde{S} -\tilde{S}d\tilde{T} -\tilde{\Phi}d\tilde{Q} -\tilde{Q}d\tilde{\Phi} - \mu dC -Cd\mu=-\tilde{S}d\tilde{T}-Cd\mu-\tilde{Q}d\tilde{\Phi}+\tilde{\Psi}d\tilde{P} - \tilde{\mathcal{P}}d\tilde{V}, \label{eq:dF2}\
	\end{equation}
and so at fixed $(\tilde{\Phi},\tilde{P},\tilde{\mathcal{V}},\mu)$, the free energy $F_4$ is stationary.  The explicit form of $F_4$ is given by
	\begin{equation}
		F_4 = \frac{\pi\tilde{P}^2(4\pi C(\tilde{\Phi},\mu)+\tilde{S})}	{\sqrt{C(\tilde{\Phi},\mu)\tilde{V}\left[(4\pi C(\tilde{\Phi},\mu)+\tilde{S})\left(\tilde{S}^2+4\pi C\tilde{S}+2\pi^2 (\tilde{P}^2+\tilde{Q}^2(\tilde{\Phi},\mu))\right)\right]}},
        \label{eq:F4explicit}
	\end{equation}
where the functions $\tilde{Q}(\tilde{\Phi},\mu)$ and $C(\tilde{\Phi},\mu)$ are obtained by solving (\ref{eq:PhiCFT}) and \eqref{eq:muCFT} using (\ref{eq:squaredE}). This  must be done numerically since the simultaneous solution to these equations entails solving a 5th-order polynomial. Note that $F_4$ is always positive.  This case will be similar to  $F_2$.
 	
	We depict the relation between $F_4$ and $\tilde{T}$   in Fig. \ref{fig:F4vsT} for various values of   $\tilde{\Phi}$ in Figs. \ref{fig:F4vsT} (a) and (b) with $\mu=1$ and $\mu=-1$, respectively;   in Fig. \ref{fig:F4vsT} (c), the plot is for various values of $\mu$ for $\tilde{\Phi}=1$. Throughout we set $\tilde{P}=0.3$ and $\tilde{V}=1$. 

If $\mu>0$ then the magnitude of $\tilde{\Phi}$ is bounded; for $\mu=1$, we find $|\tilde{\Phi}| \leq 1.452$. We see from Fig. \ref{fig:F4vsT} (a)  that there is always a zeroth-order phase transition for positive $\mu$.  The free energy curve consists of two states for  $T > T_{min}$, the value of $T$ at the cusp. Both states terminate at $F_4=0$ at two temperatures $T_1$ and $T_2$ where $T_1 \leq T_2$; these value merge when  $\tilde{\Phi}$ attains its maximal magnitude. The upper state corresponds to a low-entropy state and the lower state to a high-entropy state.
    The latter is 
     thermodynamically preferred phase   when $T_{min} < \tilde{T} < T_1$. The high-entropy state terminates at $\tilde{T} = T_1$, where   the CFT experiences a zeroth-order phase transition to the low-entropy CFT phase as the temperature increases.
	
	For $\mu=-1$ in Fig. \ref{fig:F4vsT} (b), the free energy curve again consists of two states at temperatures greater than the cusp value $T_{min}$.  For any given $\tilde{\Phi}$ the lower branch of the curve corresponds to a high-entropy state and the upper branch corresponds to a low-entropy state. There is
    a threshold value of $\tilde{\Phi}$ above which a 
    zeroth-order phase transition can always occur for every $\tilde{\Phi}$ when $\mu<0$; otherwise there is no such transition. For $\mu=-1$, a zeroth-order phase transition occurs when $2.54 \leq \tilde{\Phi} \leq 2.681$ (green, blue lines), whereas if $0 \leq \tilde{\Phi} < 2.54$ (purple, black, red lines), there is no phase transition. Negative $\tilde{\Phi}$ will give the same value as the positive one.
    
    In Fig. \ref{fig:F4vsT} (c), we compare the plots of $F_4$ versus $T$ for  different $\mu$ as $\mu=-1,-0.5,0,0.5,1$ with $\tilde{\Phi}=1$. There are two possible states for $\mu\neq 0$ and one single state for $\mu=0$ (green line). There is no phase-transition when $\mu\leq 0$ (green, black, red lines). Nonetheless, the zeroth-order phase transition occurs when $\mu>0$ (purple, blue lines).

	\subsection{Ensemble with fixed $(\tilde{\Phi},\tilde{\Psi},\tilde{V},C)$}
	\label{subsec:PhiPsiVC}

\begin{figure*}
		\centering
		\begin{tabular}{c c c}
			\includegraphics[width=.34\linewidth,valign=m,margin=-0.4cm -0.1cm]{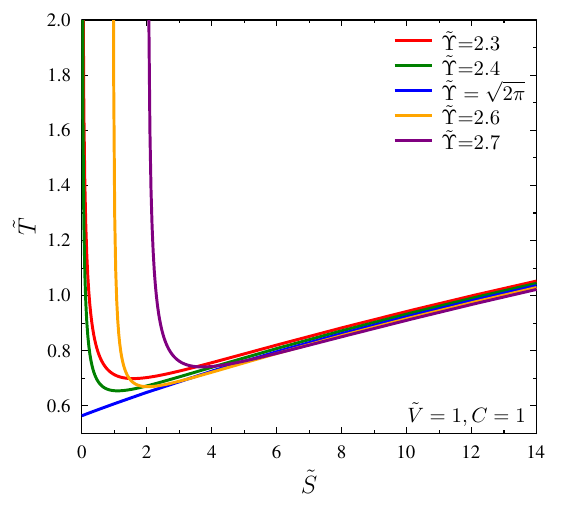} & \includegraphics[width=.34\linewidth,valign=m,margin=-0.2cm -0.1cm]{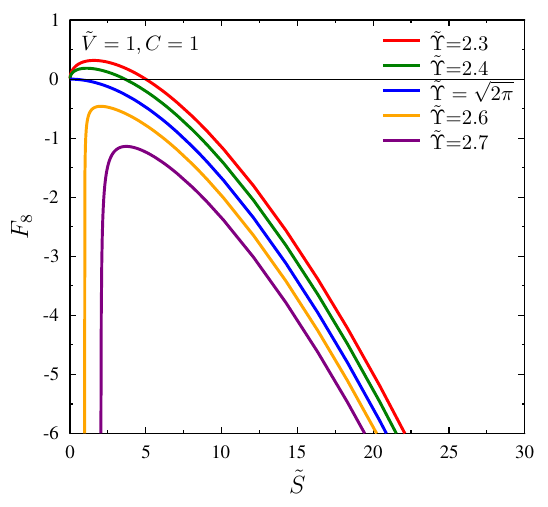} & \includegraphics[width=.34\linewidth,valign=m,margin=-0.3cm -0.1cm]{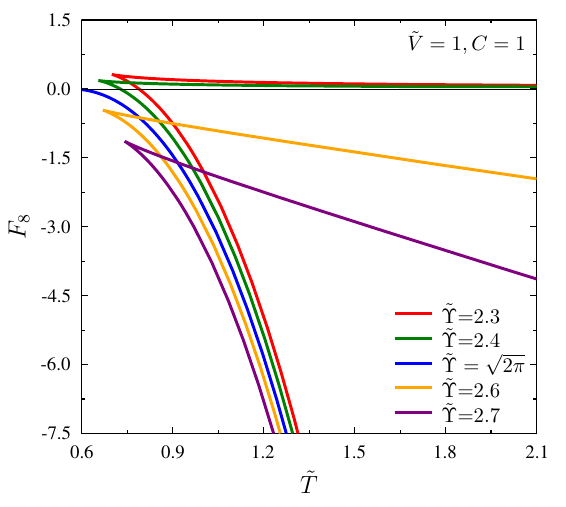}\\
			(a) & (b) & (c) \
		\end{tabular}
		\caption{(a) Plot of temperature $\tilde{T}$ vs entropy $\tilde{S}$, (b) free energy $F_8$ vs entropy $\tilde{S}$, and (c) free energy $F_8$ vs temperature $\tilde{T}$ for the ensemble at fixed $(\tilde{\Phi},\tilde{\Psi},\tilde{V},C)$. We set the parameters $C=1$ and $\tilde{V}=1$ and vary $\tilde{\Upsilon}=2.3, 2.4, \sqrt{2\pi}, 2.6, 2.7$.}
		\label{fig:F8vsT}
	\end{figure*}

    The remaining ensemble of interest  is  the one where both the electric and magnetic potentials are fixed, along with the  spatial volume $\tilde{V}$, and central charge $C$.
	The free energy of this ensemble is $F_8 = E- \tilde{T}\tilde{S}-\tilde{\Phi}\tilde{Q}-\tilde{\Psi}\tilde{P}$ as given in Eq. (\ref{eq:F8}). Its differential form is 
	\begin{equation}
		dF_8 = dE- \tilde{T}d\tilde{S} -\tilde{S}d\tilde{T} -\tilde{\Phi}d\tilde{Q} -\tilde{Q}d\tilde{\Phi}-\tilde{\Psi}d\tilde{P} -\tilde{P}d\tilde{\Psi}=-\tilde{S}d\tilde{T}-\tilde{Q}d\tilde{\Phi}-\tilde{P}d\tilde{\Psi} - \tilde{\mathcal{P}}d\tilde{V} +\mu dC, \label{eq:dF3}\
	\end{equation}
	and so, $F_8$ is stationary at fixed $(\tilde{\Phi},\tilde{\Psi},\tilde{V},C)$. Explicitly, we have
	\begin{equation}
		F_8 = \frac{16\pi^2 C^2 \tilde{S} -2\pi^2 \tilde{S}(\tilde{Q}^2(\tilde{\Phi},\tilde{\Psi})+\tilde{P}^2(\tilde{\Phi},\tilde{\Psi})) -\tilde{S}^3 }{4\pi\sqrt{C\tilde{V}\left[(4\pi C+\tilde{S})\left(\tilde{S}^2+4\pi C\tilde{S}+2\pi^2 (\tilde{P}^2(\tilde{\Phi},\tilde{\Psi})+\tilde{Q}^2(\tilde{\Phi},\tilde{\Psi}))\right)\right]}},\label{eq:F8explicit}
	\end{equation}
    where
    \begin{eqnarray}
        \tilde{P}(\tilde{\Phi},\tilde{\Psi}) &=& \pm\frac{\Psi}{\pi} \sqrt{C\tilde{V} \frac{\tilde{S}^2+4\pi C\tilde{S}}{\tilde{S}+4\pi C -2C(\tilde{\Phi}^2
        +\tilde{\Psi}^2)\tilde{V}} }, \\
        \tilde{Q}(\tilde{\Phi},\tilde{\Psi}) &=& \pm\frac{\Phi}{\pi} \sqrt{C\tilde{V} \frac{\tilde{S}^2+4\pi C\tilde{S}}{\tilde{S}+4\pi C -2C(\tilde{\Phi}^2
        +\tilde{\Psi}^2)\tilde{V}} },
    \end{eqnarray}
which are obtained by solving \eqref{eq:PhiCFT} and \eqref{eq:PsiCFT} using \eqref{eq:squaredE}. Inserting these expressions into \eqref{eq:F8explicit} yields
\begin{eqnarray}
		F_8 &=& \frac{16\pi^2 C^2 \tilde{S} -8\pi C^2\tilde{V}\tilde{S} \Upsilon^2 -\tilde{S}^3 }{4\pi\sqrt{C\tilde{V}\left[(4\pi C+\tilde{S})\left(\tilde{S}^2+4\pi C\tilde{S}-2C \tilde{V} \tilde{S}\tilde{\Upsilon}^2\right)\right]}},\label{eq:F8explicit2}
       \\
      T &=& \frac{3\tilde{S}^2 -4\tilde{S}C \left(\tilde{\Upsilon}^2\tilde{V} -4\pi    \right) -8\pi C^2\left(\tilde{\Upsilon}^2\tilde{V} -2\pi   \right) }{4\pi\sqrt{C\tilde{V}\left[(4\pi C+\tilde{S})\left(\tilde{S}^2+4\pi C\tilde{S}-2C \tilde{V} \tilde{S}\tilde{\Upsilon}^2\right)\right]}},\label{eq:Texplicit2}
	\end{eqnarray}
where $\tilde{\Upsilon}^2 = \tilde{\Phi}^2+\tilde{\Psi}^2$.  The temperature has a  minimum where $S$ satisfies
\begin{equation}\label{T8d}
    3\tilde{S}^4 -8\tilde{S}^3C \left(\tilde{\Upsilon}^2\tilde{V} -4\pi    \right)  -48\tilde{S}^2 C^2 \left(\tilde{\Upsilon}^2\tilde{V} -2\pi    \right)
    -64\pi^2 C^4\left(\tilde{\Upsilon}^2\tilde{V} -2\pi \right)^2 =0,
\end{equation}
which has at most one solution for $S>0$.  This is easily seen by applying the rule of signs -- regardless of the value of $\tilde{\Upsilon}^2\tilde{V}$, there can be at most one sign change between any of the coefficients in \eqref{T8d}. At $\tilde{S}=0$, we can have only one entropy state where $\tilde{\Upsilon}=\sqrt{2\pi/\tilde{V}}$. We want to see the relation between $\tilde{T},F_8, \tilde{S}$ when we vary $\tilde{\Upsilon}$ in spite of varying $Z$. The plots of temperature $\tilde{T}$ and free energy $F_8$ as a function of entropy $\tilde{S}$ are shown in Fig. \ref{fig:F8vsT} (a) and (b), respectively. We set $C=\tilde{V}=1$ and vary $\tilde{\Upsilon}$ as $\tilde{\Upsilon}=2.3, 2.4, \sqrt{2\pi}, 2.6, 2.7$. Only when $\tilde{\Upsilon}=\sqrt{2\pi}$, the single-state CFT is shown. Otherwise, the plots show low- and high-entropy states below and above this critical value.
    
    We display the plot of $F_8 (\tilde{T})$ parametrically in Fig. \ref{fig:F8vsT} (c). The Fig. \ref{fig:F8vsT} (c) portrays different behavior on, above and below a certain critical potential $\tilde{\Upsilon}$. On the critical electric potential $\tilde{\Phi}$ (blue line), the free energy $F_8$ is single-valued as a function of temperature with no phase transition which occurs. Above the critical potential ($\tilde{\Upsilon}=2.6, 2.7$), we obtain the free energy $F_8$ curves which consist of the upper and lower branches that meet at a cusp where the minimum temperature exists (orange and purple lines). These upper and lower branches correspond to low-entropy (dual to small black hole) and high-entropy (dual to large black hole) states, respectively.  Above the critical electric potential, we find the zeroth-order phase transition which is denoted by a finite jump in the free energy at $\tilde{T}=T_{min}$ from the thermal AdS space to high-entropy state.
    
    Below the critical potential ($\tilde{\Upsilon}=2.3, 2.4$), we also obtain the free energy $F_8$ curve which consists of the upper and lower branches that meet at a cusp where the minimum temperature exists (red and green lines) likewise to that of above critical potential. These upper and lower branches also correspond to low- and high-entropy states, respectively. Below the critical electric potential, we find different phase transition which is the first-order phase transition denoted by the switching sign of the free energy $F_8$ of the high-entropy state at $F_8=0$. This phase transition is also known as (de)confinement phase transition \cite{CongKubiznakJHEP2022}. The high-entropy or deconfined state dominates the ensemble when $F_8<0$, while the low-entropy or confined state is thermodynamically preferred when $F_8>0$. This (de)confinement phase transition is dual to the generalized Hawking-Page phase transition between large AdS black holes and the AdS spacetime with thermal radiation. The (de)confinement phase transition occurs at the temperature as follows
	\begin{equation}
		T_{F8} = \frac{\tilde{S}^2+8\pi C\tilde{S}+16\pi^2  C^2}{2\pi\sqrt{C\tilde{V}\left[(4\pi C+\tilde{S})\left(\tilde{S}^2+4\pi C\tilde{S}+2\pi^2 (\tilde{P}^2+\tilde{Q}^2)\right)\right]}},\label{eq:TF8}
	\end{equation}
	where $\tilde{Q}=\tilde{Q}(\tilde{\Phi})$ and $\tilde{P}=\tilde{P}(\tilde{\Psi})$. In terms of $\tilde{\Upsilon}$, we can write (\ref{eq:TF8}) as
    \begin{equation}
        T_{F8} = \frac{\sqrt{(\tilde{S}+4\pi C)(\tilde{S}+4\pi C - 2C\tilde{V}\tilde{\Upsilon})}}{2\pi\sqrt{C\tilde{S}\tilde{V}}}.\label{eq:TF8_2}
    \end{equation}


	\section{Thermodynamic Ensembles in the CFT for constant $\tilde{\mathcal{P}}$}\label{sec:ensemble_constantPr}
    
In this section, we will consider the remaining ensembles which are for fixed pressure. The eight ensembles at fixed pressure are
    \begin{eqnarray}
		&&\text{fixed}~(\tilde{Q},\tilde{P},\tilde{\mathcal{P}},C): ~~~ F_9 = E - \tilde{T}\tilde{S}+\tilde{\mathcal{P}}\tilde{V}=\tilde{\Phi}\tilde{Q}+\tilde{\Psi}\tilde{P} +\tilde{\mathcal{P}}\tilde{V}+\mu C, \label{eq:F9}\\
		&&\text{fixed}~(\tilde{Q},\tilde{P},\tilde{\mathcal{P}},\mu): ~~~F_{10} = E - \tilde{T}\tilde{S} +\tilde{\mathcal{P}}\tilde{V}-\mu C=\tilde{\Phi}\tilde{Q}+\tilde{\Psi}\tilde{P} +\tilde{\mathcal{P}}\tilde{V}, \label{eq:F10}\\
		&&\text{fixed}~(\tilde{\Phi},\tilde{P},\tilde{\mathcal{P}},C): ~~~F_{11} = E - \tilde{T}\tilde{S}-\tilde{\Phi}\tilde{Q} +\tilde{\mathcal{P}}\tilde{V}=\tilde{\Psi}\tilde{P}+\tilde{\mathcal{P}}\tilde{V}+\mu C, \label{eq:F11}\\
		&&\text{fixed}~(\tilde{\Phi},\tilde{P},\tilde{\mathcal{P}},\mu): ~~~ F_{12} = E - \tilde{T}\tilde{S}-\tilde{\Phi}\tilde{Q} +\tilde{\mathcal{P}}\tilde{V}-\mu C=\tilde{\Psi}\tilde{P}+\tilde{\mathcal{P}}\tilde{V}, \label{eq:F12}\\
		&&\text{fixed}~(\tilde{\Phi},\tilde{\Psi},\tilde{\mathcal{P}},\mu): ~~~ F_{13} = E - \tilde{T}\tilde{S}-\tilde{\Phi}\tilde{Q} -\tilde{\Psi}\tilde{P}+\tilde{\mathcal{P}}\tilde{V}-\mu C= \tilde{\mathcal{P}}\tilde{V}, \label{eq:F13}\\
		&&\text{fixed}~(\tilde{Q},\tilde{\Psi},\tilde{\mathcal{P}},\mu): ~~~ F_{14}= E -\tilde{T}\tilde{S}-\tilde{\Psi}\tilde{P}+\tilde{\mathcal{P}}\tilde{V}-\mu C=\tilde{\Phi}\tilde{Q}+\tilde{\mathcal{P}}\tilde{V}, \label{eq:F14}\\
		&&\text{fixed}~(\tilde{Q},\tilde{\Psi},\tilde{\mathcal{P}},C): ~~~ F_{15} = E - \tilde{T}\tilde{S}-\tilde{\Psi}\tilde{P}+\tilde{\mathcal{P}}\tilde{V}=\tilde{\Phi}\tilde{Q}+\tilde{\mathcal{P}}\tilde{V}+\mu C, \label{eq:F15}\\
		&&\text{fixed}~(\tilde{\Phi},\tilde{\Psi},\tilde{\mathcal{P}},C): ~~~ F_{16} = E - \tilde{T}\tilde{S}-\tilde{\Phi}\tilde{Q}-\tilde{\Psi}\tilde{P}+\tilde{\mathcal{P}}\tilde{V}=\tilde{\mathcal{P}}\tilde{V}+\mu C, \label{eq:F16P}\
	\end{eqnarray}
For these ensembles we need to describe the spatial volume in terms of thermodynamic pressure instead of the other way around. By employing the equation of state $E=2\tilde{\mathcal{P}}\tilde{V}$ and Eq. (\ref{eq:pCFT}), we   obtain  
	\begin{equation}
		\tilde{V}(\tilde{\mathcal{P}}) =\left(\frac{(4\pi C + \tilde{S})\left(\tilde{S}^2+2\pi^2 (\tilde{Q}^2+\tilde{P}^2) +4\pi C\tilde{S}\right)}{16\pi^2 C \tilde{\mathcal{P}}^2}\right)^{1/3}.\label{eq:Vinp}\
	\end{equation}
In  studying the free energy of each ensemble in this section, we will use this equation to define the spatial volume. Similar to the ensembles with fixed $\tilde{V}$, due to the symmetry between $\tilde{P}$ and $\tilde{Q}$ in equations \eqref{eq:squaredE} -- \eqref{eq:pCFT}, $F_{15}$ and $F_{14}$ are qualitatively similar to $F_{11}$ and $F_{12}$ respectively, and so they likewise do not warrant further analysis.

 The thermodynamic behaviour is considerably richer and more subtle than in the previous set of ensembles with fixed $\tilde{V}$. We shall consider the most interesting ensembles in some detail, with attention paid to local stability as determined by the specific heat,  given as follows
\begin{equation}
\mathcal{C} = T \left(\frac{\partial S}{\partial T}\right)_{x},
\end{equation}
where $x$ denotes the fixed quantities in the ensembles, $x\in 
\{(\tilde{Q},\tilde{P},\tilde{\mathcal{P}},\mu)$,\ $(\tilde{\Phi},\tilde{P},\tilde{\mathcal{P}},\mu),$ $(\tilde{\Phi},\tilde{\Psi},\tilde{\mathcal{P}},\mu),$ $(\tilde{Q},\tilde{P},\tilde{\mathcal{P}},C),$ $(\tilde{\Phi},\tilde{P},\tilde{\mathcal{P}},C),$  \ $(\tilde{\Phi},\tilde{\Psi},\tilde{\mathcal{P}},C)$\}.

\subsection{Ensemble with fixed $(\tilde{Q},\tilde{P},\tilde{\mathcal{P}},\mu)$}
	\label{subsec:QPPrmu}

    	\begin{figure*}
		\centering
		\begin{tabular}{c c c}
			\includegraphics[width=.35\linewidth,valign=m,margin=-0.4cm -0.1cm]{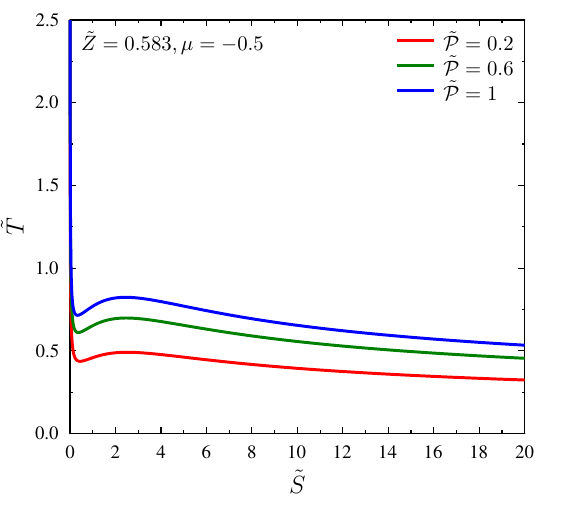} & \includegraphics[width=.35\linewidth,valign=m,margin=-0.35cm -0.1cm]{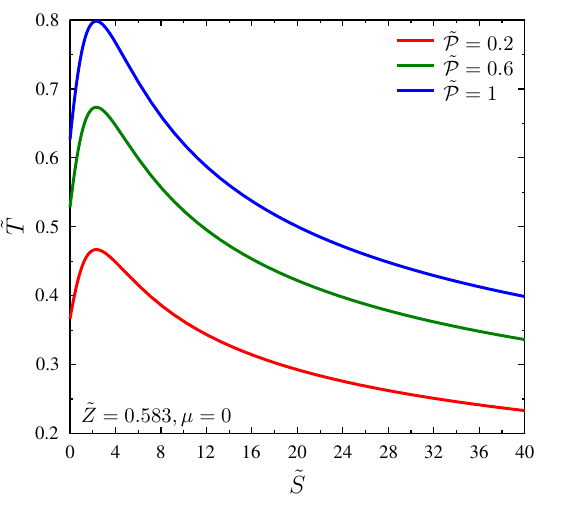} &
			\includegraphics[width=.35\linewidth,valign=m,margin=-0.35cm -0.1cm]{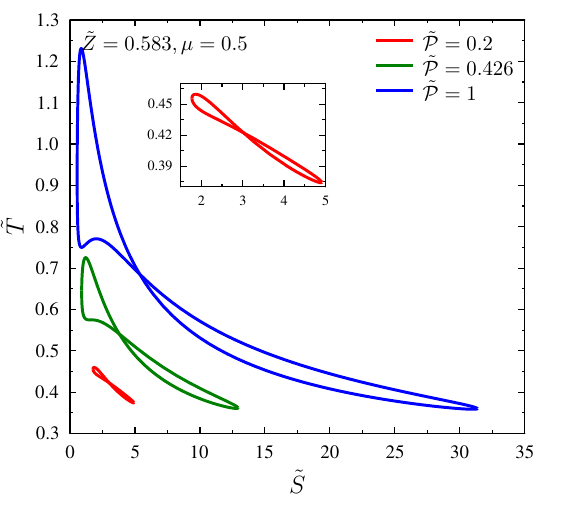} \\
			(a) & (b) & (c) \\
            \includegraphics[width=.35\linewidth,valign=m,margin=-0.4cm -0.1cm]{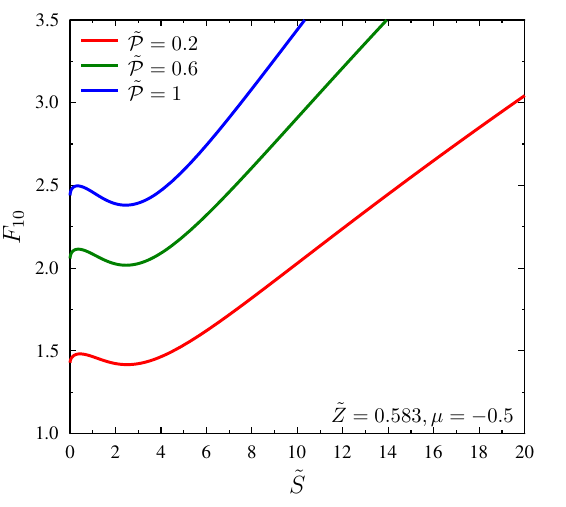} & \includegraphics[width=.35\linewidth,valign=m,margin=-0.35cm -0.1cm]{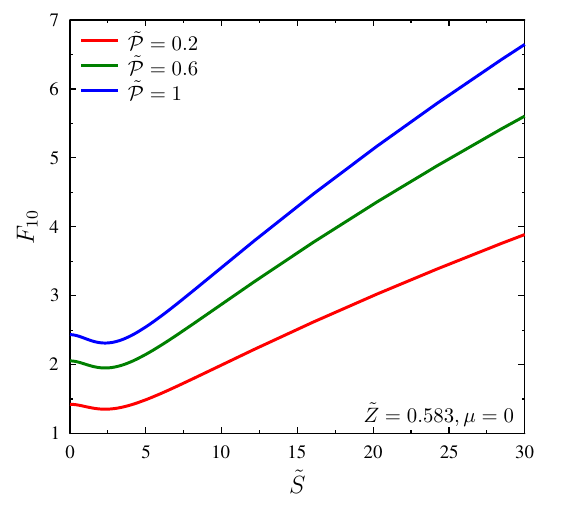} &
			\includegraphics[width=.35\linewidth,valign=m,margin=-0.35cm -0.1cm]{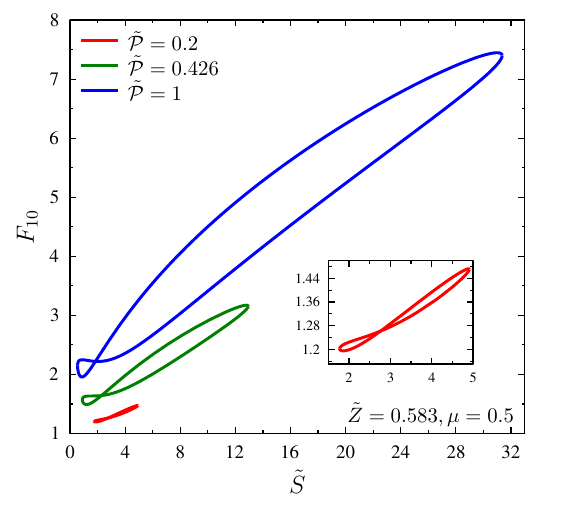} \\
			(d) & (e) & (f) \\
        	\includegraphics[width=.35\linewidth,valign=m,margin=-0.4cm -0.1cm]{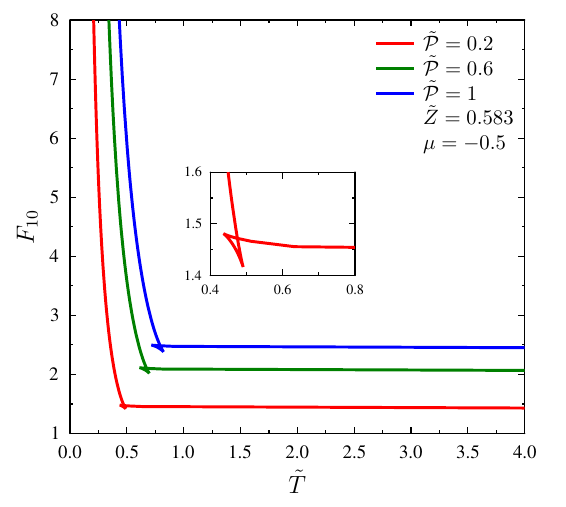} & \includegraphics[width=.35\linewidth,valign=m,margin=-0.35cm -0.1cm]{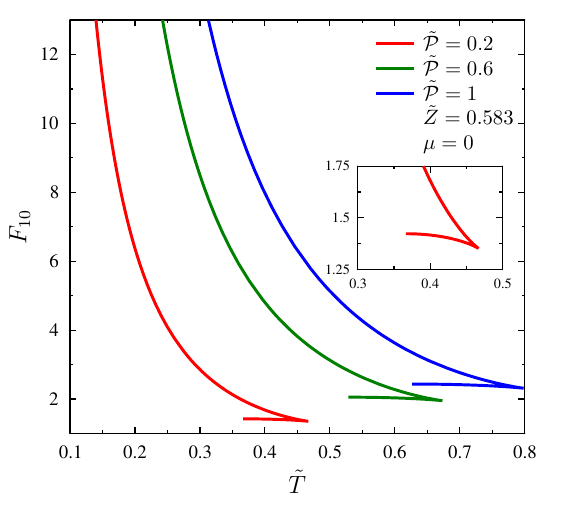} & \includegraphics[width=.35\linewidth,valign=m,margin=-0.35cm -0.1cm]{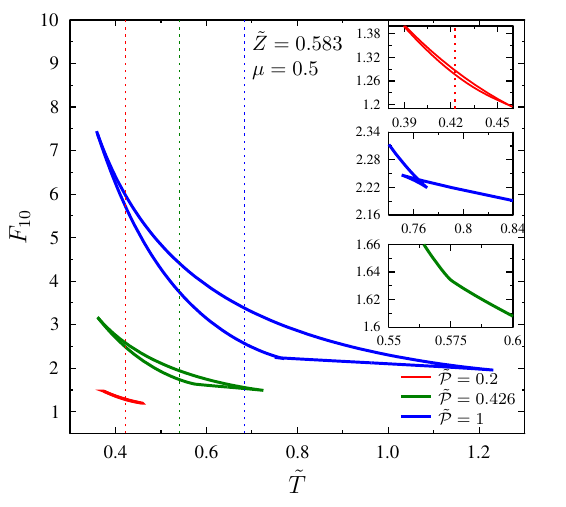} \\
			(g) & (h) & (i) 
		\end{tabular}
		\caption{\textit{Top}: Plots of  temperature $\tilde{T}$ vs entropy $\tilde{S}$ for the fixed  $(\tilde{Q},\tilde{P},\tilde{\mathcal{P}},\mu)$ ensemble for various choices of the pressure  $\tilde{\mathcal{P}}=0.2, 0.6, 1$, $\tilde{Z}=0.583$, (a) $\mu=-0.5$, (b) $\mu=0$, and (c) $\mu=0.5$.   \textit{Middle}: Plots of  free energy $F_{10}$ vs entropy $\tilde{S}$ with the same choices of parameters as in the top panels. \textit{Bottom}: Plots of  free energy $F_{10}$ vs temperature $\tilde{T}$ with the similar choices of parameters as top panels. Red curves in Figs. (g), (h) and green curve in Fig. (i) show the critical values of the pressure when the inverted swallowtails vanish.}
        
		\label{fig:F10vsST}
	\end{figure*}

	\begin{figure*}
	\centering
	\begin{tabular}{c c }
	\includegraphics[width=.4\linewidth,valign=m,margin=-0.3cm -0.1cm]{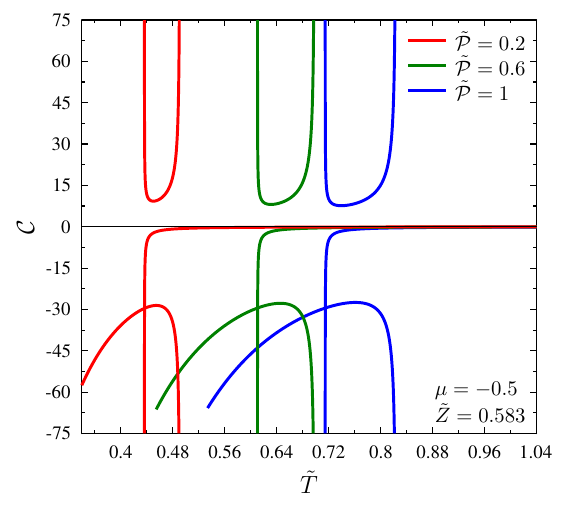} & \includegraphics[width=.4\linewidth,valign=m,margin=-0.3cm -0.1cm]{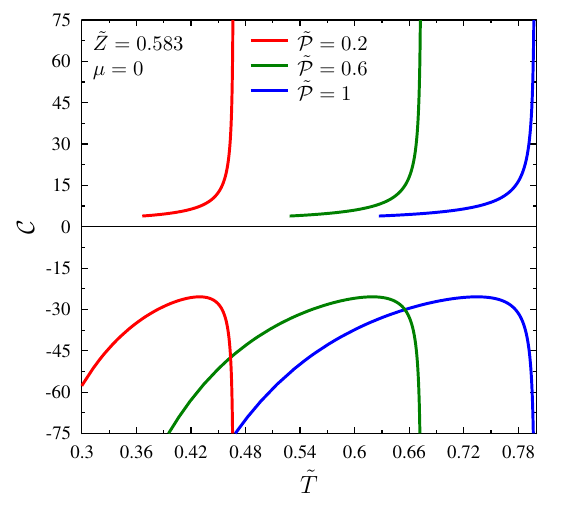} \\
    (a) & (b) \\
	\end{tabular}
    \begin{tabular}{c}
	\includegraphics[width=0.8\linewidth,valign=m,margin=-0.3cm -0.1cm]{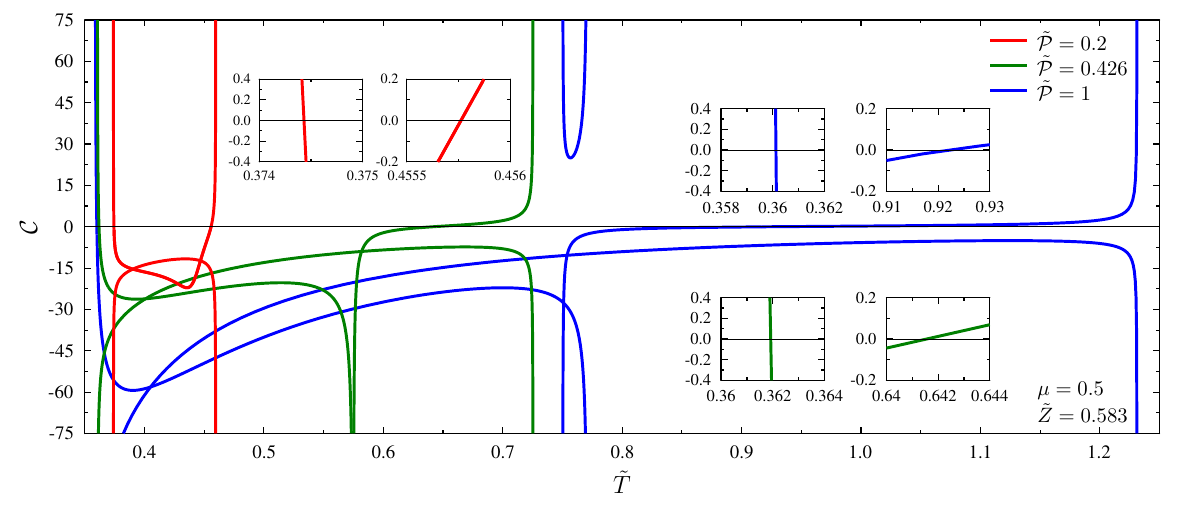} \\
    (c)  \\
	\end{tabular}
	\caption{Plots of specific heat $\mathcal{C}$ vs temperature $\tilde{T}$ for the fixed $(\tilde{Q},\tilde{P},\tilde{\mathcal{P}},\mu)$
    ensemble for (a) $\mu=-0.5$, (b) $\mu=0$, and (c) $\mu=0.5$. We generally set $\tilde{Z}=0.583$. In (a) and (b) we vary $\tilde{\mathcal{P}}$ as $\tilde{\mathcal{P}}=0.2, 0.6, 1$, and (c) $\tilde{\mathcal{P}}=0.2, 0.426,1$.
  The different colours correspond to those in Fig.~\ref{fig:F10vsST}. }
	   \label{fig:Cap10vsT}
	\end{figure*}
    
	
	We begin by considering  the ensemble with fixed electric charge $\tilde{Q}$,  magnetic charge $\tilde{P}$,   and  chemical potential $\mu$. The free energy in this ensemble is given $F_{10}$ in  Eq. (\ref{eq:F10}), which satisfies the differential relation
	\begin{equation}
		dF_{10} = dE- \tilde{T}d\tilde{S} -\tilde{S}d\tilde{T} - \mu dC -Cd\mu + \tilde{\mathcal{P}}d\tilde{V}+\tilde{V}d\tilde{\mathcal{P}} =-\tilde{S}d\tilde{T}-Cd\mu+\tilde{\Phi}d\tilde{Q}+\tilde{\Psi}d\tilde{P} + \tilde{V}d\tilde{\mathcal{P}},\label{eq:dF10}\
	\end{equation}
showing that  at fixed $(\tilde{Q},\tilde{P},\tilde{\mathcal{P}},\mu)$, the free energy $F_{10}$ is stationary. Explicitly, $F_{10}$ reads as
	\begin{equation}
		F_{10} = \frac{(4\pi C(\mu)+\tilde{S})\left(\tilde{S}^2+4\pi C(\mu)\tilde{S}+6\pi^2 (\tilde{P}^2+\tilde{Q}^2)\right)}{4\pi\sqrt{C(\mu)\tilde{V}(\tilde{\mathcal{P}})\left[(4\pi C(\mu)+\tilde{S})\left(\tilde{S}^2+4\pi C(\mu)\tilde{S}+2\pi^2 (\tilde{P}^2+\tilde{Q}^2)\right)\right]}},\label{eq:F10explicit}
	\end{equation}
	where $\tilde{V}(\tilde{\mathcal{P}})$ is given by Eq. (\ref{eq:Vinp}). By inserting this into Eq. (\ref{eq:muCFT}) we can obtain 
$C(\mu)$; however the solutions are given by the roots of an 8th-order polynomical, so we shall solve for $C(\mu)$ numerically to find  the free energy $F_{10}$  as  a function of temperature.  We note that both $F_{10}$ and $\tilde{T}$ are functions of $\tilde{Z}^2=(\tilde{P}^2+\tilde{Q}^2)$ and so we need only specify the value for $\tilde{Z}$ in obtaining the plots.	 

In the top panel of Fig. \ref{fig:F10vsST}, we display the  temperature as a function of entropy;  the plots of free energy as a function of the entropy and temperature are shown in middle and bottom panels respectively. From left to right in each row we display the three distinct values of fixed $\mu$ 
(negative, zero, and positive) for  various values of the thermodynamic pressure $\tilde{\mathcal{P}}$, with $\tilde{Z} = 0.583$. Each plot exhibits features characteristic of a globally unstable black hole with significant $\mu$-dependent features worth noting.

In the top row of Fig. \ref{fig:F10vsST}, we observe for $\mu<0$  that there is a local minimum of the temperature for small $\tilde{S}$ followed by a local maximum at larger $\tilde{S}$. For larger values of  $\tilde{S}$ the temperature  monotonically decreases to  zero, and  as 
 $\tilde{S}\to 0$ the temperature diverges. This divergent behaviour does not occur for $\mu\geq 0$. For  $\mu=0$ the temperature increases from a finite value at $\tilde{S}=0$ to a single global maximum at a finite value of $\tilde{S}$, after which it  monotonically decreases to zero at large $\tilde{S}$. For $\mu>0$ (right panel, top row)   the temperature is double-valued, with an additional curve monotonically decreasing from a global maximum at small $\tilde{S}$ to a global minimum at large $\tilde{S}$.  With increasing pressure, the maxima increase and the minima decrease, as the blue and green curves respectively illustrate.  As pressure decreases, a point of inflection occurs at a critical value of the pressure; below this value, (red curve) the temperature has only a single global minimum, as shown in the inset.

The  free energy $F_{10}$ as a function of entropy (middle row) 
exhibits behaviour analogous to that of the temperature for the corresponding values of $\mu$, except that the roles of maxima and minima are interchanged. There are two minima of the free energy when  $\mu<0$ (Fig. \ref{fig:F10vsST} (d)), there is one minimum  when  $\mu=0$ (Fig. \ref{fig:F10vsST} (e)), and at sufficiently large pressure there exist two minima when  $\mu>0$  (Fig. \ref{fig:F10vsST} (f)). Note that for $\mu>0$ the local minimum in the Fig. \ref{fig:F10vsST} (i) becomes a point of inflection at a critical pressure (green curve); below this pressure, the free energy has only a single global minimum at small $\tilde{S}$, as shown in the inset.  

 In the bottom panels, we plot the free energy $F_{10}$ as a function of temperature. Here the behaviour is somewhat more subtle, and in Fig.~\ref{fig:Cap10vsT} we plot the specific heat $\mathcal{C}$ as a function of temperature so as to discern the regions of local stability (where $\mathcal{C}>0$).

 For $\mu<0$,  Fig.~\ref{fig:F10vsST} (g), there exist two unstable   phases punctuated by a stable phase over a narrow range of temperature as indicated by  the inverted swallowtail. The  near-vertical line corresponds to an unstable  high-entropy phase whereas the
near-horizontal line corresponds to an unstable low-entropy phase; The corresponding specific heats are respectively given by the bumps and hook curves for 
$\mathcal{C}<0$ in Fig.~\ref{fig:Cap10vsT} (a).  The lower part of the inverted swallowtail corresponds to a stable phase of entropy values intermediate between these two; the specific heats are given by the corresponding U-shaped curves of $\mathcal{C}>0$ in Fig.~\ref{fig:Cap10vsT} (a). 
 The size of the inverted swallowtail grows with increasing pressure. This is very similar to the behaviour of uncharged asymptotically flat Lovelock black holes \cite{WuMannCQG2023}. In the low-temperature state the CFT is unstable, and will attain higher temperatures as the degrees of freedom radiate. Once the temperature reaches the left-hand part of the inverted swallowtail, there will be a zeroth-order phase transition to the colder end of the stable region.  As the temperature increases, the system becomes more stable.  However, once the hottest temperature of the stable region is attained, the only remaining states for the CFT are those of negative specific heat, albeit of very small value as indicated by  the near-horizontal lines at the right in the left panel, corresponding to the near-flat ends of the hook-shaped curves in  Fig.~\ref{fig:Cap10vsT} (a). 
 The inverted swallowtail is present for all
$\tilde{\cal{P}} > 0$.

 For $\mu=0$ in Fig.~\ref{fig:F10vsST} (h), there are only two thermodynamic phases for the CFT: an unstable high-entropy phase (upper curve) with $\mathcal{C}<0$ and a locally stable  low-entropy phase (lower curve) with $\mathcal{C}>0$, clearly shown as the respective bump and steep-sloped curves in  Fig.~\ref{fig:Cap10vsT} (b). The free-energy curves meet at a cusp where the temperature attains its maximal value. As with the $\mu<0$ case, at low temperatures the CFT can exist only in the unstable phase, but at sufficiently large temperature there will be a zeroth-order phase transition to the stable phase. As the temperature increases, the system becomes increasingly stable until the maximal temperature is attained.

The most interesting behaviour is for $\mu>0$, shown in Figs. \ref{fig:F10vsST} (i) and~\ref{fig:Cap10vsT} (c).  Though complicated, there are several features that stand out.
\begin{enumerate}
    \item An inverted swallowtail appears at high pressures (blue curve) and vanishes at low pressures (red curve); there is a critical pressure ($\tilde{\mathcal{P}}_c = 0.426$) at which the swallowtail just vanishes (green curve) as the pressure approaches $\tilde{\mathcal{P}}_c$ from above.
    \item The $\mu>0$ case has both  maximum and minimum temperatures,  entropies, and free energies. Consequently, free energy is a double-valued function of the temperature, resulting in the triangle-like shapes  in  Fig.~\ref{fig:F10vsST} (i).
\item    Each curve in  Fig.~\ref{fig:F10vsST} (i) has a cusp at its upper left low-temperature end and lower right high-temperature end, where the specific heat diverges, as shown in Fig.~\ref{fig:Cap10vsT} (c).
 \item The upper branches of $F_{10}$ all have negative specific heat, shown by the bump-shaped curves in  Fig.~\ref{fig:Cap10vsT} (c).    
    \item The lower branches of $F_{10}$ likewise all have negative specific heat, with two notable exceptions. Near each cusp there exists a small range of temperatures where the specific heat is positive on these branches. As temperature increases, the regions of temperature at which $\mathcal{C}$ goes from positive to negative (at low temperature) or negative to positive (at high temperature are shown in the insets of Fig.~\ref{fig:Cap10vsT} (c). For example the low-entropy branch of the green curve has positive specific heat for $\tilde{T} < 0.362$ and $\tilde{T} >0.6415$.  Hence if the system begins at some intermediate temperature where $\mathcal{C}<0$, its temperature  will increase as it radiates until it reaches the stable region at large temperature (about 0.46 for the red curve).
    \item The lower branch of $F_{10}$ (which has the inverted swallowtail for large $\tilde{\mathcal{P}}$) has higher entropy than the upper branch at low temperatures, but has lower entropy at large temperatures. The crossover temperatures where this transition takes place are indicated by the vertical dashed lines in  Fig.~\ref{fig:F10vsST} (i).    
\end{enumerate}

 In addition to these common features, there are a few idiosyncratic ones.  The lower branch of the inverted swallowtail has $\mathcal{C}>0$, indicated by the blue U-shaped curve in Fig.~\ref{fig:Cap10vsT} (c).  There are two zeroth-order phase transitions on the blue curve, at the tips of each swallowtail.  If the system is cooled from high temperature, it will proceed leftward along the lower blue branch until the right tip of the inverted swallowtail is reached, at which there will be a zeroth-order phase transition to the bottom branch of the inverted swallowtail.  As it is cooled further (or alternatively, as the system naturally radiates energy and cools, it will reach the left tip of the inverted swallowtail.  It will stay there unless it is externally cooled further; if it is cooled further, it will undergo another zeroth-order phase transition to the upper left part of the bottom branch. The swallowtail intersection also corresponds to a first-order phase transition of the unstable low and high entropy phases of the lower branch. As the critical pressure is approached from above, this phase transition becomes second-order, and the specific heat is divergent, shown in the green curve for $\tilde{T} \sim 0.57$ in  Fig.~\ref{fig:Cap10vsT} (c). There are no locally stable phases for pressures
$\tilde{\mathcal{P}} <\tilde{\mathcal{P}}_c$, shown as the red curves in Figs. \ref{fig:F10vsST} (i) and~\ref{fig:Cap10vsT} (c), except near each cusp in which there exists a small range of temperatures where the specific heat is positive as we have mentioned.

	\subsection{Ensemble with fixed $(\tilde{\Phi},\tilde{P},\tilde{\mathcal{P}},\mu)$}
	\label{subsec:PhiPPrmu}
	
\begin{figure*}
		\centering
		\begin{tabular}{c c}			\includegraphics[width=.4\linewidth,valign=m,margin=-0.4cm -0.1cm]{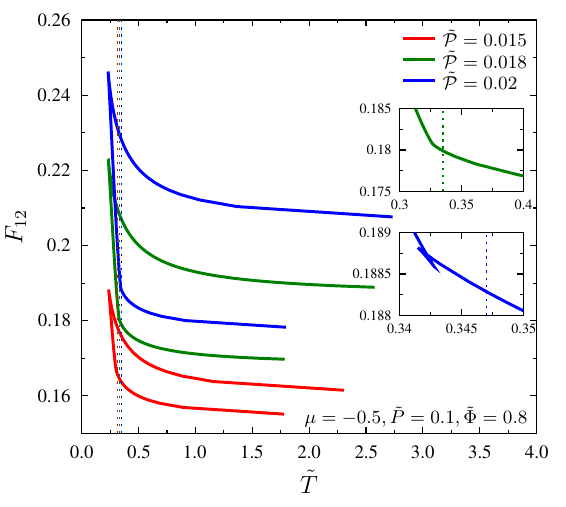} & \includegraphics[width=.4\linewidth,valign=m,margin=-0.4cm -0.1cm]{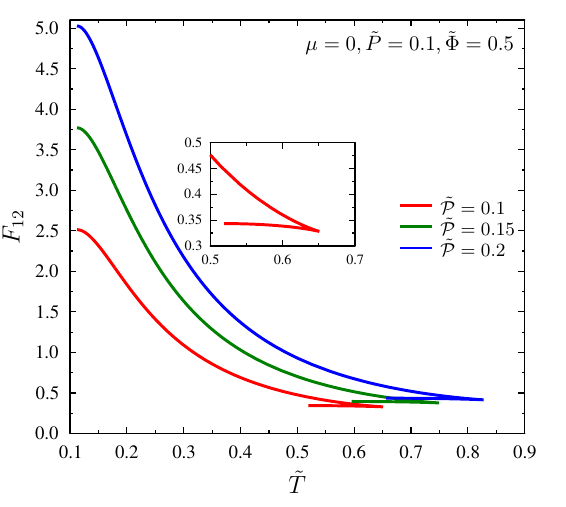}  \\ 
         (a) & (b)  \\ \includegraphics[width=.4\linewidth,valign=m,margin=-0.3cm -0.1cm]{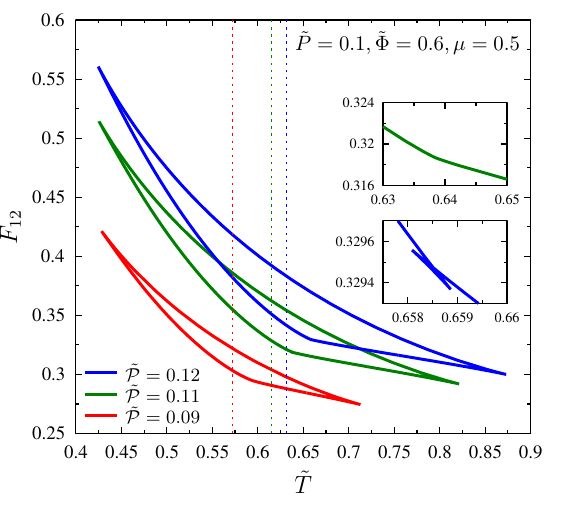} 
        & \includegraphics[width=.4\linewidth,valign=m,margin=-0.4cm -0.1cm]{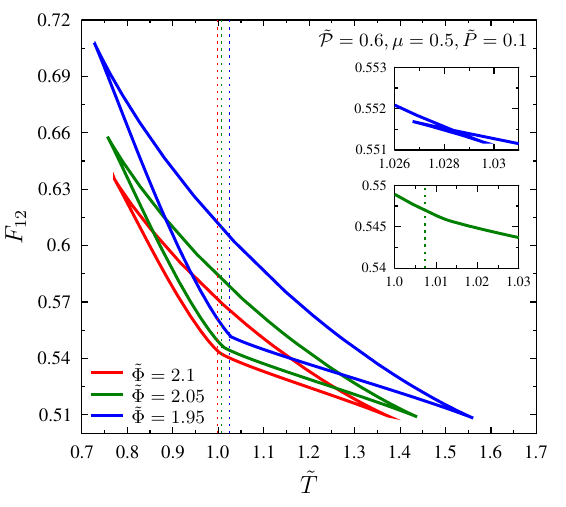}   \\
            (c)&(d)\
		\end{tabular}
		\caption{Plot of free energy $F_{12}$ vs temperature for ensemble with fixed $(\tilde{\Phi},\tilde{P},\tilde{\mathcal{P}},\mu)$ with the variation of $\tilde{\mathcal{P}}$ in (a)-(c) and variation of $\Phi$ in (d). We generally set $\tilde{P}=0.1$, and $\tilde{\Phi}=0.8, \mu=-0.5$ in (a), $\tilde{\Phi}=0.5, \mu=0$ in (b), $\tilde{\Phi}=0.6, \mu=0.5$ in (c), and $\tilde{\mathcal{P}}=0.6, \mu=0.5$. We also generally use $\tilde{P}=0.1$. The curves in (a)  terminate at large temperatures, dependent on   the range of $S$;  when $S$ is very small, the terminal temperature   will be very large. The upper branches in   (b) have higher entropy, whereas for   (a,c,d) there is a crossover indicated by the vertical  dashed lines; to the left of these lines, the lower branch has higher entropy than the upper branch.
       }
		\label{fig:F12vsT}
	\end{figure*}

\begin{figure*}
	\centering
	\begin{tabular}{c c }
	\includegraphics[width=.4\linewidth,valign=m,margin=-0.3cm -0.1cm]{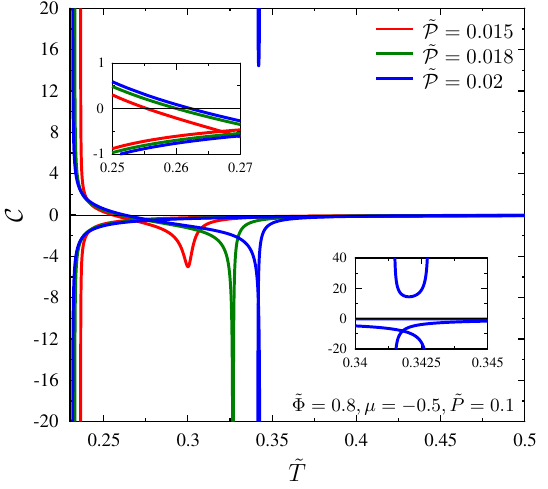} & \includegraphics[width=.4\linewidth,valign=m,margin=-0.3cm -0.1cm]{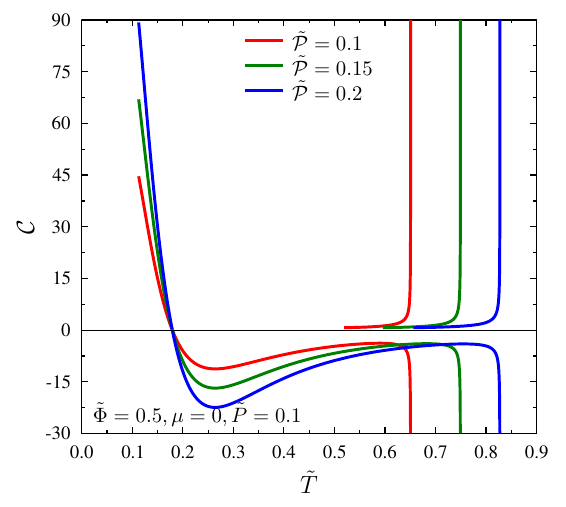} \\
    (a) & (b) \\
	\end{tabular}
    \begin{tabular}{c}
	\includegraphics[width=0.8\linewidth,valign=m,margin=-0.3cm -0.1cm]{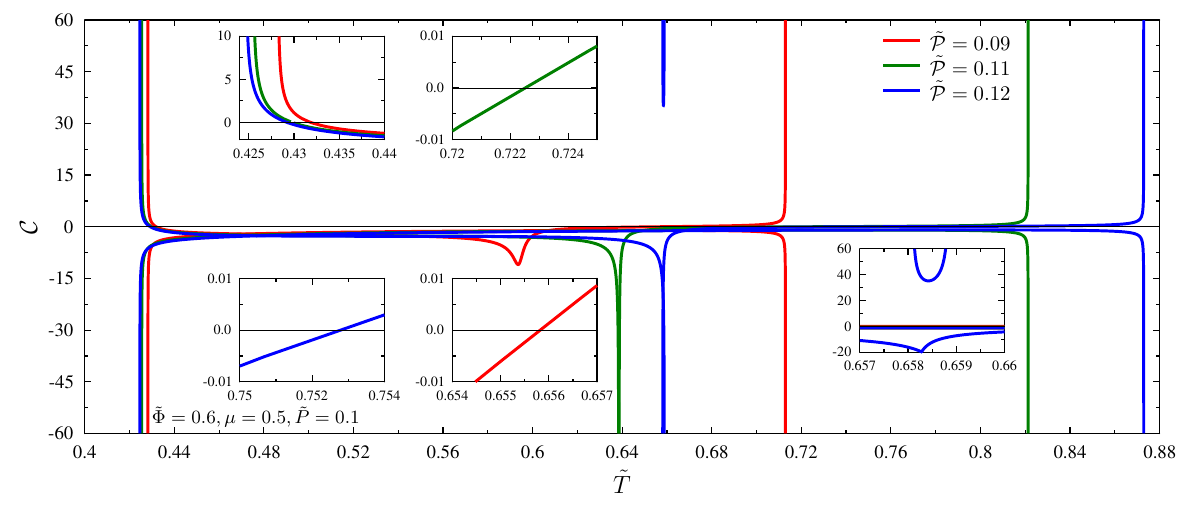} \\
    (c)  \\
	\end{tabular}
        \begin{tabular}{c}
	\includegraphics[width=0.8\linewidth,valign=m,margin=-0.3cm -0.1cm]{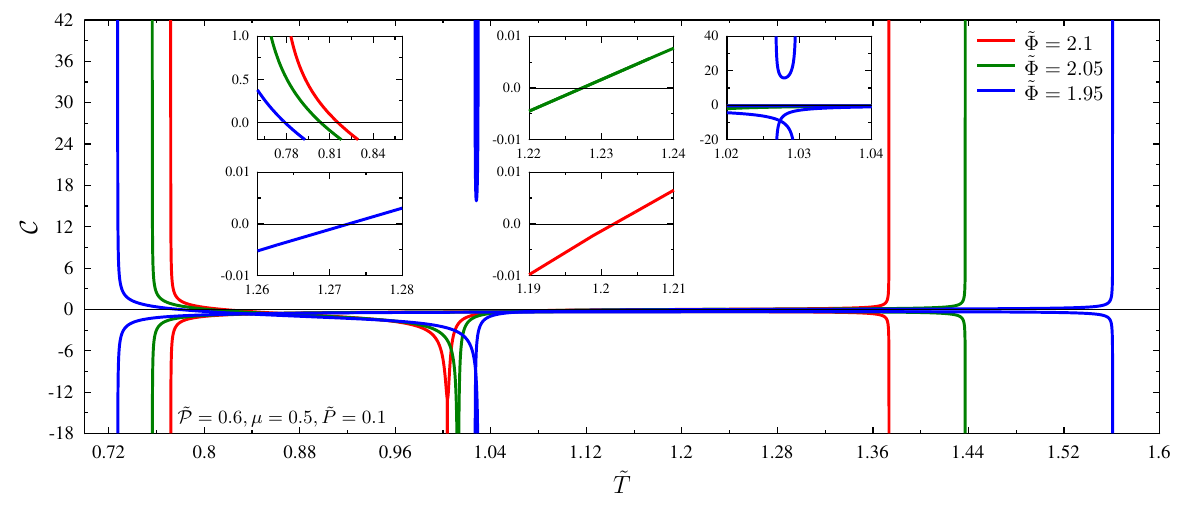} \\
    (d)  \\
	\end{tabular}
	\caption{Plot of specific heat $\mathcal{C}$ vs temperature for ensemble with  fixed $(\tilde{\Phi},\tilde{P},\tilde{\mathcal{P}},\mu)$ with the variation of $\tilde{\mathcal{P}}$ in (a)-(c) and variation of $\Phi$ in (d). We generally set $\tilde{P}=0.1$, and $\tilde{\Phi}=0.8, \mu=-0.5$ in (a), $\tilde{\Phi}=0.5, \mu=0$ in (b), $\tilde{\Phi}=0.6, \mu=0.5$ in (c), and $\tilde{\mathcal{P}}=0.6, \mu=0.5$. We also generally use $\tilde{P}=0.1$.}
	   \label{fig:Cap12vsT}
	\end{figure*}

	The ensemble with fixed $(\tilde{\Phi},\tilde{P},\tilde{\mathcal{P}},\mu)$
    has the free energy
    $F_{12}$, whose differential is
	\begin{eqnarray}
		dF_{12} &=& dE- \tilde{T}d\tilde{S} -\tilde{S}d\tilde{T} -\tilde{\Phi}d\tilde{Q} -\tilde{Q} d\tilde{\Phi}- \mu dC -Cd\mu + \tilde{\mathcal{P}}d\tilde{V}+\tilde{V}d\tilde{\mathcal{P}},\nonumber\\ &=&-\tilde{S}d\tilde{T}-Cd\mu -\tilde{Q}d\tilde{\Phi}+\tilde{\Psi}d\tilde{P} + \tilde{V}d\tilde{\mathcal{P}} \label{eq:dF12}\
	\end{eqnarray}
showing that at fixed $(\tilde{\Phi},\tilde{P},\tilde{\mathcal{P}},\mu)$, the free energy $F_{12}$ is stationary.  This ensemble is qualitatively similar to fixed $(\tilde{Q},\tilde{\Psi},\tilde{\mathcal{P}},\mu)$ ensemble with free energy $F_{14}$ (\ref{eq:F14}). The explicit form of $F_{12}$ is given by
	\begin{equation}
		F_{12} = \frac{(4\pi C(\mu)+\tilde{S})\left(\tilde{S}^2+4\pi C(\mu)\tilde{S}+2\pi^2 (3\tilde{P}^2(\tilde{\Psi})+\tilde{Q}^2(\tilde{\Phi}))\right)}{4\pi\sqrt{C(\mu)\tilde{V}(\tilde{\mathcal{P}})\left[(4\pi C(\mu)+\tilde{S})\left(\tilde{S}^2+4\pi C(\mu)\tilde{S}+2\pi^2 (\tilde{P}^2(\tilde{\Psi})+\tilde{Q}^2(\tilde{\Phi}))\right)\right]}},\label{eq:F12explicit}
	\end{equation}
	where, using \eqref{eq:Vinp}, $C(\mu)$ and $\tilde{Q}(\tilde{\Phi})$ must be computed numerically 
from  Eq. (\ref{eq:muCFT}) and (\ref{eq:PhiCFT}), since the   solutions to these equations are given by the roots of an 8th-order polynomial.  We plot the free energy $F_{12}$ (using (\ref{eq:F12explicit}) ) in Fig. \ref{fig:F12vsT} for various values of $\mu$, $\tilde{\mathcal{P}}$ and  $\tilde{\Phi}$, and the corresponding specific heats in Fig.~\ref{fig:Cap12vsT}.  For all values of $\mu$ there is a maximum and minimum temperature. Apart from this common feature, 
we again observe three qualitatively distinct types of behaviour depending on the sign of $\mu$.

 The $\mu<0$ case, shown in Figs.~\ref{fig:F12vsT} (a) and~\ref{fig:Cap12vsT} (a) is similar to the previous case, but with some important distinctions. One is that  the free energy develops a cusp at low-temperatures (instead of diverging) resulting in two branches. Another is that there is now a critical pressure 
($\tilde{\mathcal{P}}_c=0.015$ for the chosen parameters) 
above which an inverted swallowtail is present in the lower branch (blue curve) and below which it is absent (red curve); the critical case is shown by the green curve.
The upper branch (of high entropy) is locally unstable, as the closely nested bump-shaped curves in
Fig.~\ref{fig:Cap12vsT} (a) indicate, enlarged in the lower right inset. 
The lower branches all have negative specific heat, except near the cusp, where $\mathcal{C}>0$; the temperatures at which this takes place are shown in the upper left inset of Fig.~\ref{fig:Cap12vsT} (a). 
The lower portion of the inverted swallowtail has $\mathcal{C}>0$, shown as the narrow U-shaped blue curve in 
Fig.~\ref{fig:Cap12vsT} (a).  The specific heat of the lower branch diverges at the critical point (near $\tilde{T}\sim 0.335$ for the green curve); below the critical pressure   $\mathcal{C}$ has a local minimum, near $\tilde{T}\sim 0.3$ for the red curve.
 The lower branch of $F_{12}$ has higher entropy than the upper branch at low temperatures, but has lower entropy at large temperatures. The crossover temperatures where this transition takes place are indicated by the vertical dashed lines in  Fig.~\ref{fig:F12vsT} (a).

 Moving from low to high temperatures, above the critical pressure the system begins in an unstable phase for sufficiently large $\tilde{T}$.  As temperature increases it undergoes a zeroth-order phase transition to the stable phase once the left tip of the swallowtail is reached. Increasing the temperature further, eventually the hottest state in this stable phase is reached; for hotter temperatures the system can only be in an unstable phase. Alternatively, at sufficiently cold temperatures the system begins in a stable phase and will cool as it radiates. Moreover, the swallowtail intersection corresponds to a first-order transition of the unstable low- and high-entropy states of the lower branch. When the critical pressure is approached from above, the second-order phase transition occurs in which the specific heat is divergent at the kink, shown in the green curve for $\tilde{T} \sim 0.33$ in  Fig.~\ref{fig:Cap12vsT} (a).

 Note that the upper branches terminate at a higher temperature than the lower branches. Consequently at high temperatures the system will be in the unstable high-entropy phase, though it is a very weak instability since $\mathcal{C}$ is very small in magnitude, as shown in Fig.~\ref{fig:Cap12vsT} (a).

 The  $\mu=0$ case, shown in Figs.~\ref{fig:F12vsT} (b) and~\ref{fig:Cap12vsT} (b) also bears some resemblance to the previous ensemble, with a cusp at large $\tilde{T}$. The lower branch extends over a finite range of  $\tilde{T}$, as before, but the upper branch (of higher entropy) does not diverge at small $\tilde{T}$. Instead the free energy curves terminate at finite values, with the specific heat becoming positive for sufficiently small $\tilde{T}$; this is shown in Fig.~\ref{fig:Cap12vsT} (b). Curiously, the temperature at which this takes place seems to be independent of the pressure. At sufficiently large temperature, there will be a zeroth-order phase transition from unstable phase to the stable phase. Summarizing, this case has stable phases at low and high temperatures, with an unstable phase for intermediate temperatures.
 
 As before, the $\mu>0$ case, shown in
Figs.~\ref{fig:F12vsT} (c) and~\ref{fig:Cap12vsT} (c)
is the most complicated, but it does have features similar to the previous ensemble.  There is a critical point separating an inverted swallowtail structure at high pressure (blue curve; swallowtail in the inset of Fig.~\ref{fig:F12vsT} (c))
from a case with no swallowtail (red curve), the green curve illustrating the critical case.  The emergence of the inverted swallowtail denotes the existence of  two zeroth-order phase transitions, as in previous cases. Moreover, similar to $\mu<0$ case (and above the critical pressure),  moving from low to high temperatures, the system begins in an unstable phase for sufficiently large $\tilde{T}$.  As temperature increases it undergoes a zeroth-order phase transition to the stable phase once the left tip of the swallowtail is reached. Increasing the temperature further, eventually the hottest state in this stable phase is reached; for hotter temperatures the system can only be in an unstable phase.
Alternatively, at sufficiently cold temperatures the system begins in a stable phase and will cool as it radiates. At the critical pressure, the inverted swallowtail disappears denoting the emergence of the second-order phase transition.

All the features described for the previous ensemble hold in this case for $\mu>0$. The free-energy is a double-valued triangle-like curve containing two cusps where $\mathcal{C}$ diverges. Near these cusps, $\mathcal{C} > 0$, with the specific heat otherwise being negative; the transition from positive to negative (low $\tilde{T}$) or negative to positive (high $\tilde{T}$ is shown in the insets of Fig.~\ref{fig:Cap12vsT} (c). The lower branch of the inverted swallowtail has $\mathcal{C}>0$, near $\tilde{T}\sim 0.6585$ as shown in the lower right inset. The zeroth-order and first-order phase transitions occur for the pressure above the critical pressure and it will become second-order phase transition when the critical pressure is achieved. Then there is no phase transition below the critical pressure.

 We also depict the situation for various values of fixed $\tilde{\Phi}$, shown in Figs.~\ref{fig:F12vsT} (d) and~\ref{fig:Cap12vsT} (d). The qualitative features are the same as those discussed for various fixed $\tilde{\mathcal{P}}$; only the quantitative values differ. There is a critical potential $\tilde{\Phi}_c$ below which an inverted swallowtail is present (blue curve) and above which it is absent (red curve), with the green curve indicating the critical case.  All of the qualitative features discussed in the previous paragraph are present here as well.

	\subsection{Ensemble with fixed $(\tilde{\Phi},\tilde{\Psi},\tilde{\mathcal{P}},\mu)$}
	\label{subsec:PhiPsiPrmu}

Next we consider the  ensemble with both electric 
   and magnetic potentials ($\tilde{\Phi}$ and $\tilde{\Psi}$) fixed, in addition to  fixed chemical potential $\mu$.   
   In this case, the free energy is 
   $F_{13} =\tilde{\mathcal{P}}\tilde{V}$, which is  proportional to  half of the energy in CFT thermodynamics. Its differential form  is  
	\begin{eqnarray}
		dF_{13} &=& dE- \tilde{T}d\tilde{S} -\tilde{S}d\tilde{T} -\tilde{\Phi}d\tilde{Q} -\tilde{Q} d\tilde{\Phi}-\tilde{\Psi}d\tilde{P}-\tilde{P}d\tilde{\Psi}- \mu dC -Cd\mu + \tilde{\mathcal{P}}d\tilde{V}+\tilde{V}d\tilde{\mathcal{P}},\nonumber\\ &=&-\tilde{S}d\tilde{T}-Cd\mu -\tilde{Q}d\tilde{\Phi}-\tilde{\Psi}d\tilde{P} + \tilde{V}d\tilde{\mathcal{P}}. \label{eq:dF12}\
	\end{eqnarray}
showing that at fixed $(\tilde{\Phi},\tilde{\Psi},\tilde{\mathcal{P}},\mu)$, the free energy $F_{13}$ is stationary. Explicitly, we obtain 
	\begin{eqnarray}
		F_{13} &=& \frac{\sqrt{(4\pi C(\mu)+\tilde{S})\left(\tilde{S}^2+4\pi C(\mu)\tilde{S}+2\pi^2 (\tilde{P}^2(\tilde{\Psi})+\tilde{Q}^2(\tilde{\Phi}))\right)}}{4\pi\sqrt{C(\mu)\tilde{V}(\tilde{\mathcal{P}})}} \nonumber\\
        &=& \left(\frac{(4\pi C(\mu) + \tilde{S})\tilde{\mathcal{P}}\left(\tilde{S}^2+2\pi^2  (\tilde{Q}^2(\tilde{\Phi})+\tilde{P}^2(\tilde{\Psi})) +4\pi C(\mu)\tilde{S}\right)}{16\pi^2 C(\mu)   }\right)^{1/3}
        .\label{eq:F13explicit}
	\end{eqnarray}
 where $\tilde{Z}=\sqrt{\tilde{Q}^2(\tilde{\Phi}) + \tilde{P}^2(\tilde{\Psi})}$
 and $C(\mu)$ are obtained by numerically solving the equations
 \begin{eqnarray}
 \tilde{\Upsilon}^6 
     &=& 
    \frac{2^{4/3}\pi^{8/3}(4\pi C(\mu) + \tilde{S})^2 \tilde{Z}^6 \tilde{\mathcal{P}}^2}{\left(\tilde{S}^2+4\pi C(\mu)\tilde{S}+2\pi^2 \tilde{Z}^2\right)^4 C^2(\mu)}, \\
    \mu^6 &=& \frac{\left(\pi^2\tilde{S} (16 C^2(\mu) - 2\tilde{Z}^2)- \tilde{S}^3\right)^6 \tilde{\mathcal{P}}^2}{256\pi^4 (4\pi C(\mu) + \tilde{S})^2\left(\tilde{S}^2+4\pi C(\mu)\tilde{S}+2\pi^2 \tilde{Z}^2\right)^4 C^8(\mu)}.
 \end{eqnarray}
with $\tilde{\Upsilon}^2 =  \tilde{\Psi}^2 + \tilde{\Phi}^2$, and we have used (\ref{eq:PhiCFT}), (\ref{eq:PsiCFT}), and (\ref{eq:muCFT}). 
 Note that these equations imply that the free energy is a function of $\tilde{\Upsilon}$, and so it is only this quantity that needs to be specified.
 
\begin{figure*}
		\centering
		\begin{tabular}{c c}
\includegraphics[width=.45\linewidth,valign=m,margin=-0.4cm -0.1cm]{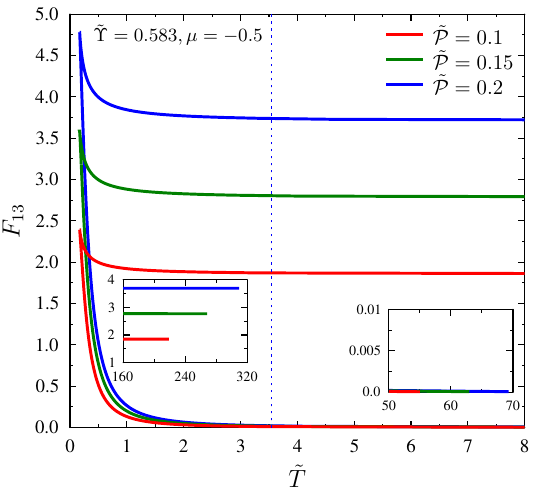} & \includegraphics[width=.45\linewidth,valign=m,margin=-0.3cm -0.1cm]{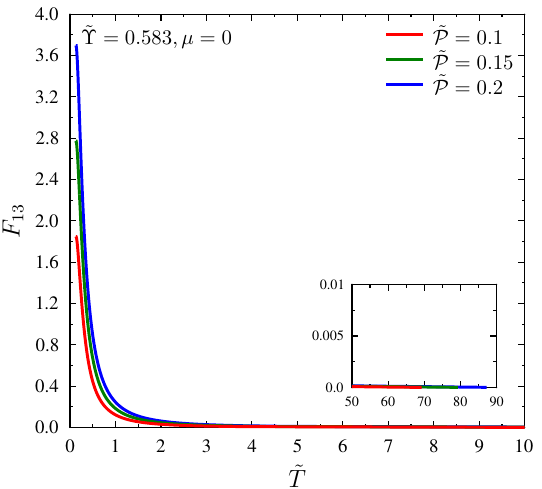} 
\\
(a) & (b) \\
\includegraphics[width=.45\linewidth,valign=m,margin=-0.3cm -0.1cm]{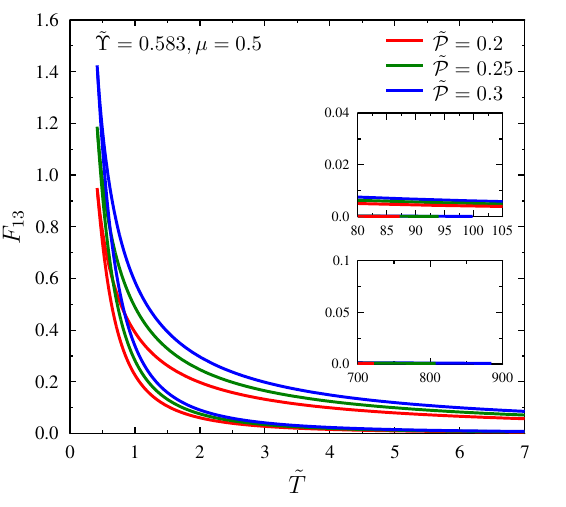}  & \includegraphics[width=.45\linewidth,valign=m,margin=-0.4cm -0.1cm]{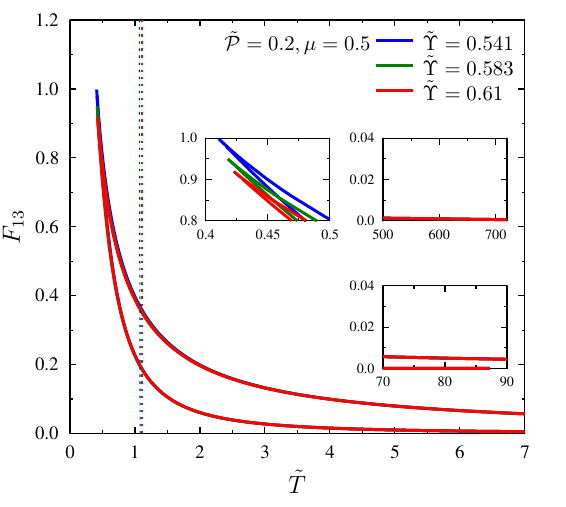}   \\
           (c) &(d) 
		\end{tabular}
		\caption{Plot of free energy $F_{13}$ vs temperature for ensemble with fixed $(\tilde{\Phi},\tilde{\Psi},\tilde{\mathcal{P}},\mu)$. In Figs. (a)-(c), we generally set $\tilde{\Upsilon}=0.583$, and (a) $\mu=-0.5$, (b) $\mu=0$, and (c) $\mu=0.5$. In Fig. (a)-(b), $\tilde{\mathcal{P}}$ is varied as $0.1, 0.15, 0.2$. In Fig. (c), $\tilde{\mathcal{P}}$ is varied as $0.2,0.25,0.3$. In Fig. (d), we set $\tilde{\mathcal{P}}=0.2, \mu=0.5$ and we vary $\Upsilon$ as $0.541, 0.583, 0.61$.}
		\label{fig:F13vsT}
	\end{figure*}
\begin{figure*}
		\centering
		\begin{tabular}{c c}
\includegraphics[width=.45\linewidth,valign=m,margin=-0.4cm -0.1cm]{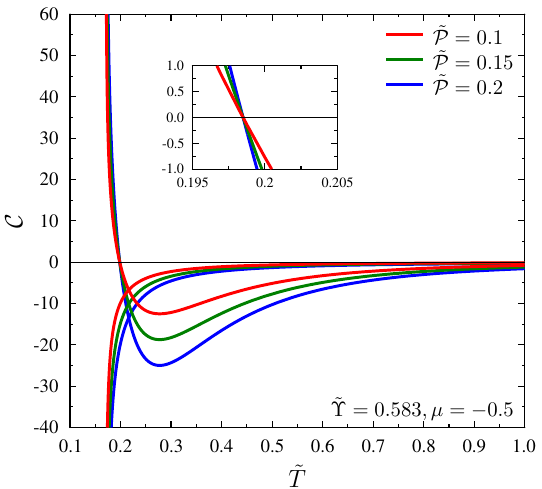} & \includegraphics[width=.45\linewidth,valign=m,margin=-0.3cm -0.1cm]{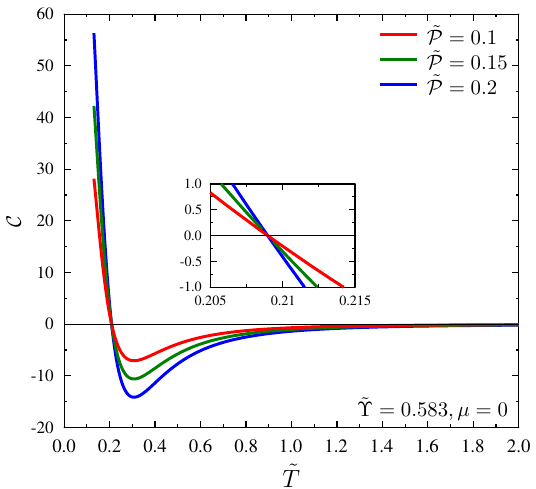} 
\\
(a) & (b) \\
\includegraphics[width=.45\linewidth,valign=m,margin=-0.3cm -0.1cm]{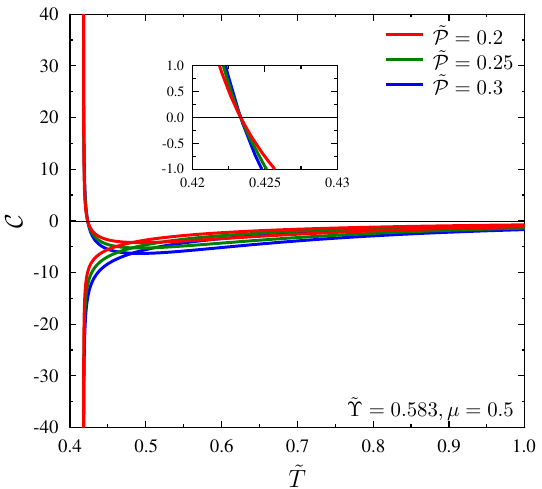}  & \includegraphics[width=.45\linewidth,valign=m,margin=-0.4cm -0.1cm]{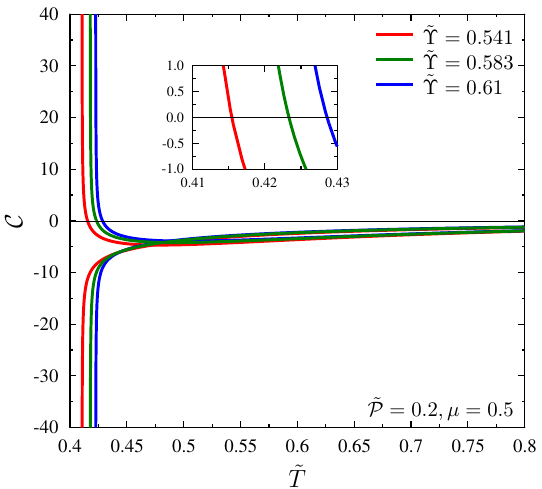}   \\
           (c) &(d) 
		\end{tabular}
		\caption{Plot of specific heat $\mathcal{C}$ vs temperature for ensemble with  fixed $(\tilde{\Phi},\tilde{\Psi},\tilde{\mathcal{P}},\mu)$. In Figs. (a)-(c), we generally set $\tilde{\Upsilon}=0.583$, and (a) $\mu=-0.5$, (b) $\mu=0$, and (c) $\mu=0.5$. In Fig. (a)-(b), $\tilde{\mathcal{P}}$ is varied as $0.1, 0.15, 0.2$. In Fig. (c), $\tilde{\mathcal{P}}$ is varied as $0.2,0.25,0.3$. In Fig. (d), we set $\tilde{\mathcal{P}}=0.2, \mu=0.5$ and we vary $\tilde{\Upsilon}$ as $0.541, 0.583, 0.61$.}
		\label{fig:Cap13vsT}
	\end{figure*}

 We  plot  the free energy $F_{13}$ as a function of the temperature $\tilde{T}$ in Fig. \ref{fig:F13vsT}, with the corresponding specific heats in
Fig.~\ref{fig:Cap13vsT}.  The dashed lines in Fig. \ref{fig:F13vsT} show the temperature crossover. The lower branch of $F_{13}$ has higher entropy (left of the dashed line) than the upper branch at low temperatures, but has lower entropy at large temperatures. As with the previous two ensembles, we observe distinct behaviour depending on the sign of $\mu$, with the case containing features similar to those of the previous two ensembles.

 However there are a number of noteworthy distinctions. One is that  we do not observe any inverted swallowtails,  unlike the previous two ensembles. The second is that there is only a single cusp present in $F_{13}$ if $\mu\neq 0$, and no cusp at all if $\mu=0$. For all values of $\mu$ there are minimum and maximum temperatures. The upper branches of $F_{13}$ are unstable for all values of $\tilde{T}$ for $\mu\neq 0$.  The lower branches, and the $\mu=0$ case are unstable for sufficiently large $\tilde{T}$, but have a region of stability ($\mathcal{C}>0$) present at low temperatures, as shown in Fig.~\ref{fig:Cap13vsT}. For $\mu>0$ the   features are similar if we consider various fixed  $\tilde{\mathcal{P}}$ or fixed $\tilde{\Upsilon}$, as a comparison of Figs.~\ref{fig:F13vsT}  (c,d) and~\ref{fig:Cap13vsT} (c,d) indicate. 

 For $\mu<0$ the upper branches terminate at a higher temperature than the lower branches, as in the previous ensemble. Thus  the system will be in the unstable high-entropy phase at high temperatures; again this is a very weak instability since $\mathcal{C}$ is very small in magnitude. As the system cools, the CFT will eventually experience a zeroth-order phase transition to the low-entropy state \cite{CongKubiznakJHEP2022,AhmedCongJHEP2023}.  Similarly for $\mu>0$, the higher-entropy branch truncates at temperatures higher (lower inset on Fig. \ref{fig:F13vsT} (c)) than the low-entropy branch, and so the higher branch is the thermodynamically preferred state. This state is also weakly unstable since $\mathcal{C}$ is very small in magnitude. Also for fixed $\tilde{\Upsilon}$, the upper branches (upper right inset on Fig. \ref{fig:F13vsT} (d)) terminate at larger $\tilde{T}$ than the lower branches, similar to the other cases.

	\subsection{Other ensembles at fixed $\tilde{\mathcal{P}}$}
	\label{subsec:other}

The remaining ensembles   are those with
fixed central charge $C$, namely $F_9$ (\ref{eq:F9}), $F_{11}$  (\ref{eq:F11}) (symmetric to $F_{15}$ (\ref{eq:F15})), and $F_{16}$ (\ref{eq:F16P}). 
      Unlike the ensembles with fixed  chemical potential $\mu$, these ensembles do not exhibit any interesting phase transitions. 
       The explicit forms of these free energies are given by
	\begin{eqnarray}
		F_{9} &=& \frac{2C(\mu)\tilde{S}^2 + 8\pi C^2 \tilde{S} + \pi \tilde{S}(\tilde{P}^2+\tilde{Q}^2) + 6\pi^2 C (\tilde{P}^2+\tilde{Q}^2) }{\sqrt{C\tilde{V}(\tilde{\mathcal{P}})\left[(4\pi C+\tilde{S})\left(\tilde{S}^2+4\pi C\tilde{S}+2\pi^2 (\tilde{P}^2+\tilde{Q}^2)\right)\right]}},\label{eq:F9explicit}\\
		F_{11} &=& \frac{2C\tilde{S}^2 + 8\pi C^2 \tilde{S} + \pi \tilde{S}\tilde{P}^2+ 2\pi^2 C (3\tilde{P}^2+\tilde{Q}^2(\tilde{\Phi})) }{\sqrt{C\tilde{V}(\tilde{\mathcal{P}})\left[(4\pi C+\tilde{S})\left(\tilde{S}^2+4\pi C\tilde{S}+2\pi^2 (\tilde{P}^2+\tilde{Q}^2(\tilde{\Phi}))\right)\right]}},\label{eq:F11explicit}\\
		F_{16} &=& \frac{2\sqrt{C}\left(\tilde{S}^2+4\pi C\tilde{S}+\pi^2 (\tilde{P}^2(\tilde{\Psi})+\tilde{Q}^2(\tilde{\Phi}))\right) }{\sqrt{\tilde{V}(\tilde{\mathcal{P}})\left[(4\pi C+\tilde{S})\left(\tilde{S}^2+4\pi C\tilde{S}+2\pi^2 (\tilde{P}^2(\tilde{\Psi})+\tilde{Q}^2(\tilde{\Phi}))\right)\right]}}.\label{eq:F16explicit}\
	\end{eqnarray}
with $\tilde{V}(\tilde{\mathcal{P}})$   given by Eq. (\ref{eq:Vinp}).  
The functions $\tilde{Q}(\tilde{\Phi})$ and $\tilde{P}(\tilde{\Psi})$ are  obtained  from Eqs. (\ref{eq:PhiCFT}) and (\ref{eq:PsiCFT}).  We note that $F_{9}$ is a function of $\tilde{Z}$ and $F_{16}$ is a function of $\tilde{\Upsilon}$.

\begin{figure}
		\centering
		\begin{tabular}{c c c}
			\includegraphics[width=.34\linewidth,valign=m,margin=-0.4cm -0.1cm]{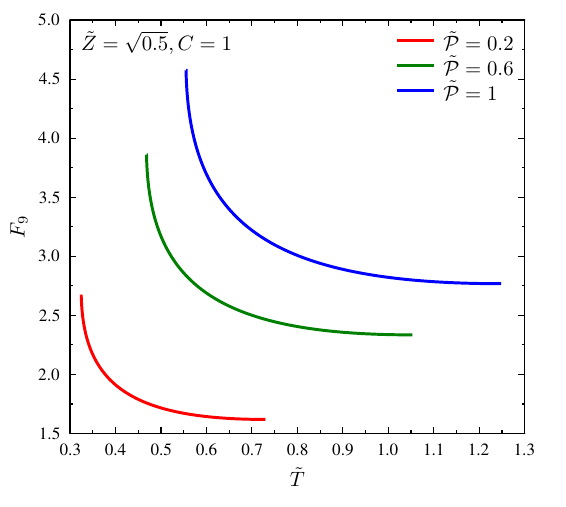} & \includegraphics[width=.34\linewidth,valign=m,margin=-0.2cm -0.1cm]{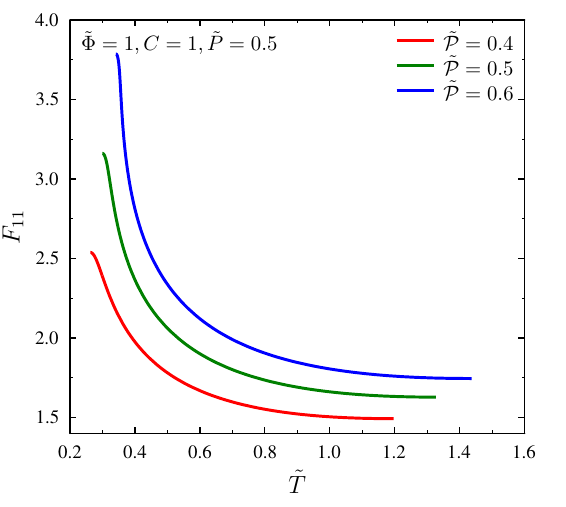} & \includegraphics[width=.34\linewidth,valign=m,margin=-0.3cm -0.1cm]{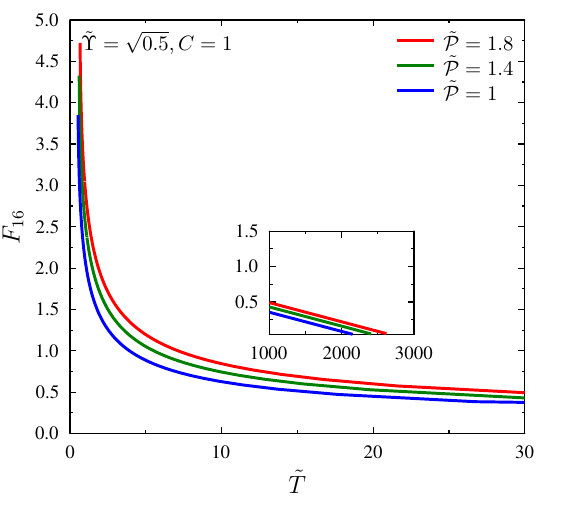}\\
			(a) & (b) & (c) \
		\end{tabular}
		\caption{(a) Plot of free energy $F_9$ vs temperature $\tilde{T}$ for the ensemble at fixed $(\tilde{Q},\tilde{P},\tilde{\mathcal{P}},C)$ where we set $\tilde{Z}=0.5, C=1$ and vary $\tilde{\mathcal{P}}=0.2, 0.6, 1$. (b) Plot of free energy $F_{11}$ vs temperature $\tilde{T}$ for the ensemble at fixed $(\tilde{\Phi},\tilde{P},\tilde{\mathcal{P}},C)$ where we set $\tilde{\Phi}=1,\tilde{P}=0.5,C=1$ and vary $\tilde{\mathcal{P}}=0.4, 0.5, 0.6$. (c) Plot of free energy $F_{16}$ vs temperature $\tilde{T}$ for the ensemble at fixed $(\tilde{\Phi},\tilde{\Psi},\tilde{\mathcal{P}},C)$ where we set $\tilde{\Upsilon}=\sqrt{0.5},C=1$ and vary $\tilde{\mathcal{P}}=1, 1.4, 1.8$.  In   (c), at very large $\tilde{T}$ or $\tilde{S}\rightarrow 0$, the free energy $F_{16}$   asymptotically approaches  zero.
        }
		\label{fig:F9_11_16}
	\end{figure}

\begin{figure}
		\centering
		\begin{tabular}{c c c}
			\includegraphics[width=.34\linewidth,valign=m,margin=-0.4cm -0.1cm]{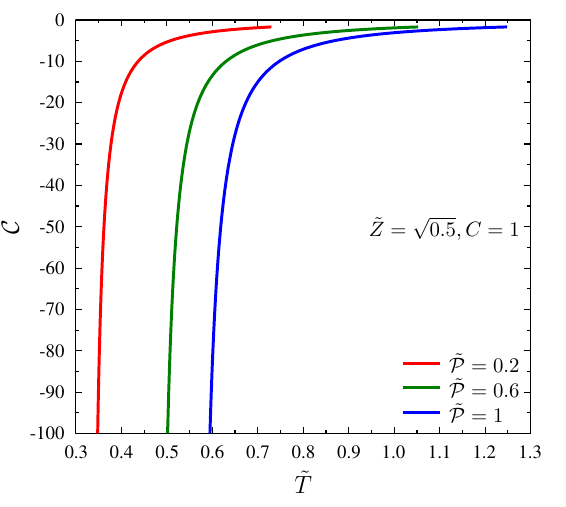} & \includegraphics[width=.34\linewidth,valign=m,margin=-0.2cm -0.1cm]{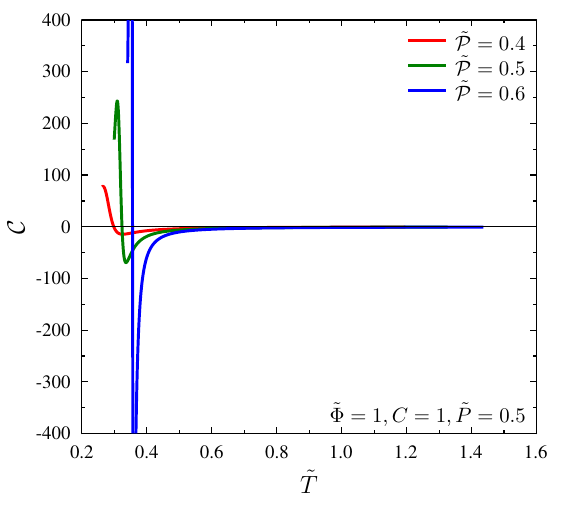} & \includegraphics[width=.34\linewidth,valign=m,margin=-0.3cm -0.1cm]{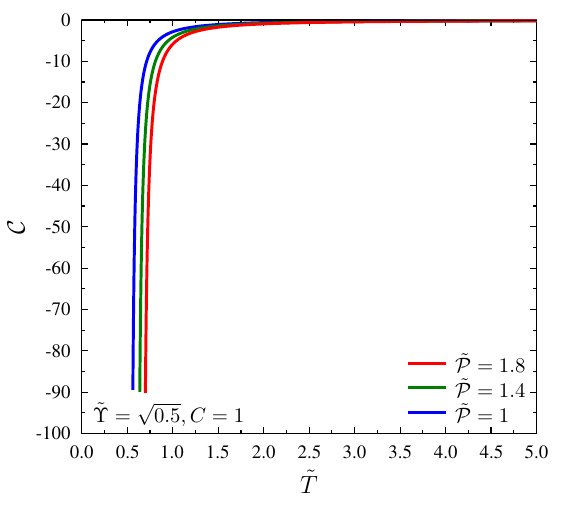}\\
			(a) & (b) & (c) \
		\end{tabular}
		\caption{(a) Plot of specific heat $\mathcal{C}$ vs temperature $\tilde{T}$ for the ensemble at fixed $(\tilde{Q},\tilde{P},\tilde{\mathcal{P}},C)$ where we set $\tilde{Z}=0.5, C=1$ and vary $\tilde{\mathcal{P}}=0.2, 0.6, 1$. (b) Plot of specific heat $\mathcal{C}$ vs temperature $\tilde{T}$ for the ensemble at fixed $(\tilde{\Phi},\tilde{P},\tilde{\mathcal{P}},C)$ where we set $\tilde{\Phi}=1,\tilde{P}=0.5,C=1$ and vary $\tilde{\mathcal{P}}=0.4, 0.5, 0.6$. (c) Plot of specific heat $\mathcal{C}$ vs temperature $\tilde{T}$ for the ensemble at fixed $(\tilde{\Phi},\tilde{\Psi},\tilde{\mathcal{P}},C)$ where we set $\tilde{\Upsilon}=\sqrt{0.5},C=1$ and vary $\tilde{\mathcal{P}}=1, 1.4, 1.8$. 
        }
		\label{fig:Cap9_11_16}
	\end{figure}

We display the plots of these free energies against temperature $\tilde{T}$  in Fig. \ref{fig:F9_11_16}. For any given  parameter choices, we find that only a single phase for $F_9$ and $F_{16}$ while there is a small range of distict phase at small temperature for $F_{11}$ (also for $F_{15}$). Furthermore, the free energy $F_{16}$ goes asymptotically to zero for very small entropy $\tilde{S}$ which is related to very large CFT temperature $\tilde{T}$.  The  specific heat plots for these ensembles are shown in Fig. \ref{fig:Cap9_11_16}. The ensembles with fixed $(\tilde{Q},\tilde{P},\tilde{\mathcal{P}},C)$ and $(\tilde{\Phi},\tilde{\Psi},\tilde{\mathcal{P}},C)$ are unstable as shown in Fig. \ref{fig:Cap9_11_16} (c). Interestingly, for the ensemble with fixed $(\tilde{\Phi},\tilde{P},\tilde{\mathcal{P}},C)$ (also for fixed $(\tilde{Q},\tilde{\Psi},\tilde{\mathcal{P}},C)$ ensemble), for the given pressures, it is stable at small temperatures and   unstable at large temperatures, as shown in Fig. \ref{fig:Cap9_11_16} (b).
	
\section{Summary}\label{sec:summary}

We have systematically investigated the holographic thermodynamics of   static dyonic dilaton AdS black holes. The metric is  obtained by setting the spin parameter to be zero for dyonic Kerr-Sen-AdS black holes, which are an exact solution of Einstein-Maxwell-Dilaton-Axion theory. The static black hole has a root at $r_0=p^2/m$ and another root at a larger value of $r$.  However, when $r_0=p^2/m$,  $\Sigma (r_0)=0$, and the spacetime is singular,  which also corresponds to  divergent gauge and dilaton fields. Since physical spacetime terminates at $r_0$, the inner horizon $r < r_0$ is physically non-existent. The metric function $\Delta(r)$ possesses only one single physical root located at a larger radius $r > r_0$, which defines the event horizon of the black hole. Because the curvature singularity is located exactly where the inner root of $\Delta(r)$ would be, the inner and outer horizons can never be brought into coincidence. Therefore, the dyonic dilaton AdS black hole cannot possess an extremal counterpart.

We first presented explicit mass/energy formulas for the extended black hole thermodynamics. By leveraging the holographic dictionary in an extended thermodynamic framework where Newton’s gravitational constant $G$ and AdS length $l$ are treated as a dynamical variable for the dyonic dilaton AdS black hole, we mapped the bulk parameters to their dual boundary CFT variables. This allowed us to explore the thermodynamic phase space featuring sixteen distinct ensembles defined under conditions of either fixed boundary volume $\tilde{V}$ or fixed boundary pressure $\tilde{\mathcal{P}}$. However, due to the symmetry between the electric charge $\tilde{Q}$ and magnetic charge $\tilde{P}$, as well as their potentials $\tilde{\Phi}$ and $\tilde{\Psi}$, several ensembles share qualitatively identical behaviors. This symmetry narrows the parameter space down to ten distinct thermodynamic classes split evenly between fixed boundary volume $\tilde{V}$ and fixed boundary pressure $\tilde{\mathcal{P}}$ frameworks. The detailed results and phase properties extracted from every single one of these 16 ensembles are summarized below.

\subsection{Fixed Spatial Volume $\tilde{V}$ Ensembles}

The fixed $(\tilde{Q}, \tilde{P}, \tilde{V}, C)$ ensemble with Helmholtz free energy $F_1$ (\ref{eq:F1}) serves as the baseline for fixed-boundary volume systems. The temperature behavior mirrors the standard U-shaped Schwarzschild-AdS black hole's curve which behaves like $(\tilde{T} \sim 1/\sqrt{\tilde{S}} + \sqrt{\tilde{S}})$. On the other hand, the temperature increases for small and large entropy. It features a sharp bifurcation point (cusp) at a minimum temperature $T_{\text{min}}$ that separates an unstable small black hole branch from a stable large black hole branch. However, since the electromagnetic charges are held fixed, there is not a Hawking-Page phase transition.

Then we considered fixed $(\tilde{Q}, \tilde{P}, \tilde{V}, \mu)$ ensemble with free energy $F_2$ (\ref{eq:F2}) by holding the chemical potential fixed instead of the central charge. The system yields positive free energy curves, which strictly forbid the emergence of first-order phase transitions. However when $\mu > 0$, the temperature curve is double-valued and exhibits a minimum threshold leading to a zeroth-order phase transition from a low-entropy state to a high-entropy state.

The ensemble with fixed $(\tilde{\Phi}, \tilde{P}, \tilde{V}, C)$ is linked symmetrically by electromagnetic duality with the ensemble with fixed $(\tilde{Q}, \tilde{\Psi}, \tilde{V}, C)$. Both ensembles correspond with the free energy $F_3$ (\ref{eq:F3}) and $F_7$ (\ref{eq:F7}), respectively. These ensembles allow the CFT to freely discharge its electric or magnetic sectors because its corresponding gauge potential is fixed. Despite this freedom, the system is prevented from completely transitioning into a pure thermal AdS space because one of the electromagnetic charge parameters remains conserved and fixed. A small parameter space window of $\tilde{\Phi}$ (or $\Psi$) exists where \(T_{\text{min}}\) occurs at negative values of entropy.

The ensemble with fixed $(\tilde{\Phi}, \tilde{P}, \tilde{V}, \mu)$ which is symmetric to the ensemble with fixed $(\tilde{Q}, \tilde{\Psi}, \tilde{V}, \mu)$. These ensembles keep both gauge and chemical potentials fixed which correspond with the free energy $F_4$ (\ref{eq:F4})and $F_6$ (\ref{eq:F6}), respectively. When $\mu > 0$, the magnitude of the electric potential $\tilde{\Phi }$ (or magnetic potential $\tilde{\Psi}$) is bounded. The system exhibits a distinct zeroth-order phase transition where a high-entropy state terminates at a critical temperature $T_{1}$, jumping directly into a low-entropy state as heating occurs to $T_2$. For $\mu < 0$, this zeroth-order phase transition vanishes unless the gauge potential exceeds a specific threshold.

The fixed $(\tilde{\Phi}, \tilde{\Psi}, \tilde{V}, C)$ ensemble with $F_{8}$ holds both the electric and magnetic potentials fixed alongside the central charge. The phase behavior depends on a critical combined potential $\tilde{\Upsilon} = \sqrt{\tilde{\Phi}^2 + \tilde{\Psi}^2}$. At the exact critical value $\tilde{\Upsilon}_c = \sqrt{2\pi/\tilde{V}}$, the system has a single-valued free energy and completely stable. Above $\tilde{\Upsilon }_{c}$, a zeroth-order phase transition emerges with a finite free energy jump at $T_{\text{min}}$ from thermal AdS space to high-entropy state. Below $\tilde{\Upsilon }_{c}$, it produces a first-order (de)confinement phase transition where the system changes sign at $F_8 = 0$, successfully mapping the dual representation of a bulk Hawking-Page transition between large AdS black hole and AdS spacetime with thermal radiation.

The last ensemble in fixed spatial volume groups is the fixed $(\tilde{\Phi}, \tilde{\Psi}, \tilde{V}, \mu)$ ensemble which corrresponds with $F_{5}$ (\ref{eq:F5}). Because all intensive conjugate variables are held constant simultaneously, the underlying Euler scaling laws dictate that the thermodynamic potential identically vanishes $F_5 = 0$. Consequently, this configuration represents a trivial unconstrained boundary system that does not warrant thermodynamic stability analysis.

\subsection{Fixed Pressure $\tilde{\mathcal{P}}$ Ensembles}

The first ensemble in the fixed pressure ensemble group is the ensemble with fixed $(\tilde{Q}, \tilde{P}, \tilde{\mathcal{P}}, \mu)$ which corresponds with the free energy $F_{10}$ (\ref{eq:F10}). This ensemble is globally unstable but it features significant $\mu-$dependent characteristics. When $\mu < 0$, the free energy curve develops a characteristic inverted swallowtail architecture bounded by two unstable phases and a stable phase $(\mathcal{C} > 0)$. At lower temperatures, a zeroth-order phase transition occurs from the cold end of the inverted swallowtail to the stable branch. For $\mu = 0$, the system contains only two thermodynamic phases—an unstable high-entropy phase $\mathcal{C}<0$ and a locally stable low-entropy phase $\mathcal{C}>0$ meeting at a maximum temperature. For $\mu > 0$, the phase structure features both local minimum and maximum temperatures, generating double-valued triangle-like loops for $\tilde{\mathcal{P}}>\tilde{\mathcal{P}}_c$ where for the given parameters, the critical pressure is $\tilde{\mathcal{P}}_c=0.426$. 
When $\tilde{\mathcal{P}}>\tilde{\mathcal{P}}_c$, it has been observed that the zeroth- and first-order phase transition occurs simultaneously. At the critical pressure, the inverted swallowtail disappears and second-order phase transition emerges. There is no phase transition occurs below the critical pressure.

For the ensemble with fixed $(\tilde{\Phi}, \tilde{P}, \tilde{\mathcal{P}}, \mu)$ which is symmetric to the ensemble with fixed $(\tilde{Q}, \tilde{\Psi}, \tilde{\mathcal{P}}, \mu)$, they are governed by free energy $F_{12}$ and $F_{14}$, respectively. These ensemble exhibit behavior highly analogous to the fixed $(\tilde{Q}, \tilde{P}, \tilde{\mathcal{P}}, \mu)$ ensemble. The main distinctions are that for $\mu < 0$, the inverted swallowtail is present only above a distinct critical pressure $\tilde{\mathcal{P}}_c$ denoting the presence of zeroth-order and first-order phase transitions while at the critical pressure, there occurs only second-order phase transition. For $\mu=0$, the unstable phase separates two distinct stable phases and the zeroth-order phase transition occurs from unstable phase to the stable phase. For $\mu > 0$, above the critical potential $\tilde{\Phi }_{c}$ (or $\tilde{\Psi}_{c}$) dictates the appearance of the inverted swallowtail denoting the presence of similar phase transitions as $\mu>0$ in the previous ensemble.

In the ensemble with fixed $(\tilde{\Phi}, \tilde{\Psi}, \tilde{\mathcal{P}}, \mu)$, we observed the absence of inverted swallowtails portrayed by free energy $F_{13} = \tilde{\mathcal{P}}\tilde{V}$. A single cusp characterizes the system when $\mu \neq 0$, and no cusps appear when $\mu = 0$. Stable configurations ($\mathcal{C} > 0$) exist strictly at low temperatures. There occurs only a zeroth-phase order transition for $\mu>0$.

The remaining ensembles are $(\tilde{Q},\tilde{P},\tilde{\mathcal{P}},C)$, $(\tilde{\Phi},\tilde{P},\tilde{\mathcal{P}},C)$, $(\tilde{\Phi},\tilde{\Psi},\tilde{\mathcal{P}},C)$ which correspond to free energies $F_9, F_{11}$, and $F_{16}$, respectively. The ensemble with $(\tilde{\Phi},\tilde{P},\tilde{\mathcal{P}},C)$ is symmetric to the ensemble with fixed $(\tilde{Q},\tilde{\Psi},\tilde{\mathcal{P}},C)$. All these ensembles do not exhibit any interesting phase transition. The ensembles which correspond to  $F_9$ and $F_{16}$ are unstable observed from the specific heat plots while the ensemble $F_{11}$ (or $F_{15}$) is stable only in small temperature regime.

Generally, comparing the results from this dyonic dilaton AdS black hole to the charged AdS black hole in Einstein-Maxwell system \cite{CongKubiznakJHEP2022}, we observe that the phase transition that occurred in each ensemble (for fixed spatial volume only; fixed pressure ensembles have not been studied) are quite different. For instance, in the ensemble with Helmholtz free energy, we do not observe any phase transition whereas for a charged AdS black hole,  first- and second-order phase transitions occur. Evidently the  presence of  dilaton and axion fields is the main factor for the interesting behaviors we observe. One cannot obtain the charged AdS black holes from dyonic dilaton AdS black hole because of the distinct interdependence of the electromagnetic, dilaton, and axion charges.

It is worth noting that in the charged AdS black hole, the fixed pressure ensembles do not exhibit any critical behavior   \cite{CongKubiznakJHEP2022}. In contrast, our results show that the fixed pressure ensembles possess richer structures and exhibit distinct phase transitions, especially due to the dependence on the  chemical potential $\mu$. Furthermore, for the rotating    charged AdS black hole   some interesting   phase transition behaviour has been observed  \cite{AhmedCongJHEP2023}. For future endeavors, we would like to consider the rotating version of the dyonic dilaton AdS black hole to see the role of angular momentum in the phase space of holographic thermodynamics. 

	\vspace*{8mm}

    \section*{Acknowledgements}
    M.F.A.R.S. is supported by the Second Century Fund (C2F) and C2F research abroad scholarship, Chulalongkorn University, Thailand. R.B.M. research is supported in part by the Natural Sciences and Engineering Research Council of Canada.  R.B.M. thanks Wan Cong for helpful discussions.

	%
    \appendix
    \section{Derivation of first law of the CFT thermodynamics}\label{sec:appA}
    Here we will derive the first law of CFT thermodynamics from the first law of bulk thermodynamics. The following relations are useful for the derivation,
    \begin{eqnarray}
    \frac{\kappa}{8\pi G}dA &=& TdS + TS \frac{dG}{G},\label{eq:f1}\\
    -\frac{V}{8\pi G} d\Lambda &=& Vd\mathcal{P} + V\mathcal{P}\frac{dG}{G} =  -(M-2TS-\Phi Q-\Psi P)\left(\frac{d\mathcal{P}}{2\mathcal{P}}+\frac{d G}{2G}\right)\nonumber\\
    &=&(M-2TS-\Phi Q-\Psi P)\frac{d l}{l} \nonumber\\
    &=&2(M-TS-\Phi Q-\Psi P) \frac{d l}{l} +(\Phi Q + \Psi P -M)\frac{d l}{l}, \label{eq:f2}\\
    dC &=& 2C\frac{dl}{l}-C\frac{dG}{G} =\left(\frac{2l}{4G} dl -\frac{l^2}{4G}\frac{dG}{G} \right),\label{eq:f3}\\
    \tilde{\Phi}d\tilde{Q}&=&\frac{\Phi \sqrt{G}}{\omega l}d\left(\frac{Ql}{\sqrt{G}}\right)=\frac{\Phi \sqrt{G}}{\omega l}\left(\frac{Q}{\sqrt{G}}dl+\frac{l}{\sqrt{G}}dQ-\frac{Ql}{2\sqrt{G}}\frac{dG}{G}\right), \label{eq:f4}\\
    \tilde{\mathcal{P}}d\tilde{V}&=& \frac{E}{2\tilde{V}}d\tilde{V}=\frac{M}{\omega l}dl.\label{eq:f5}\
    \end{eqnarray}
    We start from Eq. (\ref{eq:firstlawvariedG}),
    \begin{eqnarray}
		dM &=& \frac{\kappa}{8\pi G}dA + \Phi dQ +\Psi dP - \frac{V}{8\pi G} d\Lambda -\left(M -\frac{1}{2}\Phi Q -\frac{1}{2}\Psi P\right)\frac{dG}{G}.\nonumber\
    \end{eqnarray}
    We then write the term containing surface gravity using (\ref{eq:f1}) and cosmological pressure using (\ref{eq:f2}), such that we have
    \begin{eqnarray}
      dM  &=& \left(TdS + TS \frac{dG}{G}\right) -\left(M -\frac{1}{2}\Phi Q -\frac{1}{2}\Psi P\right)\frac{dG}{G}+ \Phi dQ +\Psi dP,\nonumber\\
        &+&2(M-TS-\Phi Q-\Psi P) \frac{d l}{l} +(\Phi Q + \Psi P -M)\frac{d l}{l},\nonumber\\
        &=&TdS +\Phi dQ + \Psi d P +(\Phi Q + \Psi P -M)\frac{d l}{l}-\frac{1}{2}(\Phi Q+\Psi P) \frac{d G}{G}
        \nonumber\\
        &+&\frac{(M-TS-\Phi Q-\Psi P)}{C} \frac{2Cd l}{l}- \frac{(M-TS-\Phi Q-\Psi P)}{C} \frac{Cd G}{G}.\nonumber\
        \end{eqnarray}
    We then divide all terms by $\omega$ and write as follows
       \begin{eqnarray}
        \frac{dM}{\omega}&=&\frac{T}{\omega}dS + \frac{\Phi \sqrt{G}}{\omega l}\left(\frac{Q}{\sqrt{G}}dl+\frac{l}{\sqrt{G}}dQ-\frac{Ql}{2\sqrt{G}}\frac{dG}{G}\right)\nonumber\\
        &+&\frac{\Psi \sqrt{G}}{\omega l}\left(\frac{P}{\sqrt{G}}dl+\frac{l}{\sqrt{G}}dP-\frac{Pl}{2\sqrt{G}}\frac{dG}{G}\right) \nonumber\\
        &+& \frac{(M-TS-\Phi Q-\Psi P)}{\omega C} \left(\frac{2l}{4G} dl -\frac{l^2}{4G}\frac{dG}{G} \right) -\frac{M}{\omega l}dl .\
	\end{eqnarray}
    By using (\ref{eq:f3})-(\ref{eq:f5}), and the AdS/CFT dictionaries (\ref{eq:AdSCFTdictionary}), we can obtain the first law of CFT thermodynamics (\ref{eq:CFTfirstlaw}),
    \begin{equation}
		dE = \tilde{T}d\tilde{S}+\tilde{\Phi}d\tilde{Q}+\tilde{\Psi}d\tilde{P}+\mu dC - \tilde{\mathcal{P}}d\tilde{V},\nonumber\
	\end{equation}
    where we define
    \begin{equation}
        \mu = \frac{M -TS -\Phi Q -\Psi P}{\omega C},
    \end{equation}
    and thus the Euler equation is given by Eq. (\ref{eq:EulerEq}).

\end{document}